\documentclass[11pt,a4paper]{article}
\usepackage[T1]{fontenc}
\usepackage{lmodern}
\usepackage{amsmath,amssymb,amsthm,mathtools}
\usepackage[a4paper,margin=1.03in,includefoot]{geometry}
\usepackage[colorlinks=true,linkcolor=blue,citecolor=blue,urlcolor=blue]{hyperref}
\newtheorem{theorem}{Theorem}[section]
\newtheorem{lemma}[theorem]{Lemma}
\newtheorem{proposition}[theorem]{Proposition}
\newtheorem{corollary}[theorem]{Corollary}
\theoremstyle{remark}

\DeclareMathOperator{\tr}{tr}
\DeclareMathOperator{\Stab}{Stab}
\DeclareMathOperator{\rad}{rad}
\DeclareMathOperator{\spanf}{span}
\newcommand{\F}{\mathbb F}
\newcommand{\E}{\mathbb E}
\newcommand{\Prob}{\mathbb P}
\newcommand{\one}{\mathbf 1}
\newcommand{\ip}[2]{\langle #1,#2\rangle}
\title{Stabilizer Learning under Global Depolarizing Noise\\with Polynomial Copies and \(2^t\) Time}
\author{Aryeh Lev Zabokritskiy (Yohananov)\\
{\small Department of Computer Science}\\
{\small Tel-Hai University of Kiryat Shmona in the Galilee}\\
{\small MIGAL---Galilee Research Institute}\\
{\small ORCID: \href{https://orcid.org/0000-0003-3151-6192}{0000-0003-3151-6192}}}
\hypersetup{
 pdftitle={Stabilizer Learning under Global Depolarizing Noise with Polynomial Copies and 2\textasciicircum t Time},
 pdfauthor={Aryeh Lev Zabokritskiy (Yohananov)},
 pdflang={en},
 pdfdisplaydoctitle=true
}
\date{}
\begin{document}
\maketitle

\begin{abstract}
We study learning the stabilizer structure of an unknown pure \(n\)-qubit state observed through global depolarizing noise of unknown strength, with a fixed promised upper bound below one, using destructive two-copy Bell measurements. Let \(0\le t\le n\) bound the stabilizer nullity, let \(\varepsilon>0\) be the accuracy, and let \(0<\delta<1\) be the failure tolerance. With probability at least \(1-\delta\), a learner must return a commuting unsigned Pauli group containing every stabilizer of the pure state; averaging conjugation by that group must decrease the purity of the observed noisy state by at most \(\varepsilon\). Tang and Steckmann gave such a learner with polynomially many copies and \(2^t\) times polynomial time when the noise strength is below a threshold of at most about \(0.16\), which tends to zero as \(t/n\to1\). They asked whether other algorithms with these resources can exceed this threshold. We answer this question affirmatively for every fixed bound on global depolarizing noise below one and every \(0\le t\le n\), including states with only one promised independent stabilizer. Our learner uses polynomially many copies and \(2^t\) times polynomial classical running time. All resource bounds hold on every execution. Their polynomial factors depend on \(n,1/\varepsilon,\log(1/\delta)\), with degrees and constants depending on the fixed noise promise.
\end{abstract}

\section{Introduction}
Learning a quantum state's stabilizer structure identifies Pauli symmetries that can compress its description. For pure states, Bell sampling determines the stabilizer group of an unknown stabilizer state up to signs, and single-copy Pauli measurements then fix the signs \cite{Mon}. Bell sampling also supports structure learning with polynomially many copies and polynomial time \cite{HG,GIKL}, and provides a primitive for tomography with additional measurements when the stabilizer nullity is small \cite{HG,GIKL}. Learning from noisy copies calls for guarantees that relate the observed mixed state to the stabilizer structure of the underlying pure state \cite{TS}.

For an \(n\)-qubit pure state whose stabilizer nullity is at most \(t\), Tang and Steckmann \cite[Sections~II.A and~III.B]{TS} formulate this problem under Pauli channels using destructive two-copy Bell measurements. A learner must return a commuting unsigned Pauli group containing the ideal stabilizers, with a bound \(\varepsilon\) on the purity lost by twirling the noisy state over that group. For global depolarization of strength \(\lambda<1\), their top-down learner uses \(4^t\) times polynomial copies and time, with polynomial dependence on \(1/(1-\lambda)\) \cite[Theorem~II.4]{TS}. For the same noise, their up-down learner achieves polynomially many copies and \(2^t\) times polynomial time under a dimension-dependent noise condition when \(n-t\gg1\) \cite[Theorem~VIII.4]{TS}. Its leading-order threshold is \(1-[(n+t)/(2n)]^{1/4}\), approaching \(1-2^{-1/4}\approx0.159\) as \(t/n\to0\). They ask whether other finders, the parts of a learner that propose candidate groups for verification, can exceed that threshold, and suggest that additional single-copy Pauli measurements may lead to a \(2^t\) algorithm above it \cite[Section~X]{TS}. With running time exponential in \(n\), polynomially many copies already suffice, by testing all \(4^n\) Pauli labels \cite[Section~X]{TS}; the threshold question concerns \(2^t\) time.

We answer this threshold question for every fixed bound on global depolarizing noise below one, throughout the range \(0\le t\le n\). Bell measurements alone suffice for polynomial copy complexity and polynomial working storage, with \(2^t\) times polynomial classical time. At fixed accuracy and confidence, for \(t=\omega(\log n)\), our guarantee improves on the top-down bound, reducing the exponential time factor from \(4^t\) to \(2^t\) and replacing exponential copy complexity by a polynomial. Its polynomial degrees and constants depend on the fixed noise promise and are very large even for the promise \(\lambda\le1/2\), whereas the top-down bound depends polynomially on \(1/(1-\lambda)\). Compared with the up-down guarantee, ours keeps the \(2^t\) time factor and extends polynomial copy complexity to every fixed noise bound below one. It also covers \(n-t=o(n)\), where the up-down threshold tends to zero, as well as \(n-t=1\) and arbitrary pure states at \(t=n\), where the condition \(n-t\gg1\) fails. Under the hypotheses of \cite[Theorems~II.2 and~VIII.5]{TS}, which add to the up-down noise condition a lower bound of order \(\lambda t/(n-t)\) on \(\varepsilon\), Tang and Steckmann also achieve polynomially many copies and polynomial time, a stronger time bound than \(2^t\) times a polynomial.

Mixed-state learning also arises under exact stabilizer promises. Grewal et al. \cite[Appendix~B]{GIKL} extend their learning guarantees to mixed states with many unit-magnitude Pauli expectations, and Cho and Kim \cite{CK} study single-copy learning under this promise. Hinsche, Eisert, and Carrasco \cite[Section~II.D]{HEC} distinguish such strong stabilizers, which their learner requires, from invariance under conjugation. Under global depolarization with positive strength, every nonidentity Pauli expectation is multiplied by \(1-\lambda\), so these exact promises fail, while for \(\lambda<1\) the Pauli conjugation symmetries of the observed state are exactly the ideal stabilizers.

Arunachalam, Bravyi, Dutt, and Yoder \cite[v2, Section~6.1, Theorem~9]{ABDY} learn globally depolarized quadratic phase states, which have stabilizer nullity zero, using additional measurements for the remaining phase information. Chen, Gong, Ye, and Zhang \cite{CGYZ} give agnostic tomography for states of stabilizer dimension \(n-t\): from copies of an arbitrary state they output a state whose fidelity with it is close to the best fidelity achieved by such states; for a fixed fidelity error, their running time is polynomial in \(n\), \(2^t\) and the inverse of that best fidelity. Closeness in fidelity does not imply stabilizer containment, since nearby states can have different stabilizer groups. Our objective is unsigned stabilizer containment with bounded additive purity loss for arbitrary pure states of bounded nullity. The output may include additional commuting directions within that loss budget.

Section~\ref{sec:main-results} defines the model, states the main theorem and outlines its proof. Section~\ref{sec:bell-tools} develops Bell identities and observed-state verification, and shows that, for one-qubit inputs, a learner using only standard Bell measurements with failure probability at most \(1/4\) needs \(\Omega((1-\lambda)^{-4})\) copies in the worst case as the noise approaches complete depolarization. Section~\ref{sec:linear-gap} treats a fixed positive relative dimension gap \((n-t)/n\). For smaller gaps, Section~\ref{sec:moment-framework} splits the input states into two overlapping cases, which are handled in Sections~\ref{sec:low-moment} and~\ref{sec:high-moment}. Section~\ref{sec:assembly} combines them into the complete learner, and Section~\ref{sec:conclusion} discusses the scope of the result and further questions.

\section{Model, main result and proof outline}\label{sec:main-results}\label{sec:framework}\label{mem:sec:all-noise}
The learning task is to recover every ideal stabilizer while controlling the purity lost by the output group. We first specify the observations and the required output, then state the resource guarantee. The proof outline explains how a list of candidate commuting spaces leads to the final verified output.

\subsection{Model and notation}\label{sec:learning-model}
We work in the label space \(V\coloneqq \F_2^{2n}\). For \(g=(x,z)\), let
\(\sigma_g\coloneqq i^{x\cdot z}X^xZ^z\), with the exponent in this representative computed as an integer. The binary symplectic form and the parity of the number of \(Y\) factors are
\[
 \ip{(x,z)}{(x',z')}\coloneqq x\cdot z'+z\cdot x',
 \qquad \pi_Y(x,z)\coloneqq x\cdot z\pmod2.
\]
The Paulis \(\sigma_g\) and \(\sigma_h\) commute exactly when \(\ip gh=0\). For a subspace \(W\le V\), write
\(W^\perp\coloneqq \{g:\ip{g}{w}=0\text{ for all }w\in W\}\) for its symplectic complement.
The subspace \(W\) is isotropic when \(W\subseteq W^\perp\), and its radical is
\(\rad(W)\coloneqq W\cap W^\perp\). Every radical is isotropic, and every isotropic subspace has dimension at most \(n\).
Since \(\pi_Y(g+h)=\pi_Y(g)+\pi_Y(h)+\ip gh\), the parity \(\pi_Y\) is linear on every isotropic subspace.

Write \(\psi\coloneqq |\psi\rangle\langle\psi|\) for a pure \(n\)-qubit state, and let \(S\coloneqq \Stab(\psi)\) be its unsigned Pauli stabilizer group: a label belongs to \(S\) when its Pauli expectation has absolute value one. Its stabilizer nullity is \(\nu\coloneqq n-\dim S\), and \(\rho\coloneqq(1-\lambda)\psi+\lambda I/2^n\) denotes its global depolarization with strength \(0\le\lambda\le1\). For a commuting unsigned Pauli group, that is, an isotropic subspace \(H\le V\), use the twirl and additive purity loss of \cite{TS},
\[
 \mathcal T_H(\rho)\coloneqq \frac1{|H|}\sum_{g\in H}\sigma_g\rho\sigma_g,
 \qquad
 \Delta\gamma(H)\coloneqq \tr(\rho^2)-\tr\bigl(\mathcal T_H(\rho)^2\bigr).
\]
Here \(\sigma_g\) is a Hermitian Pauli representative; the twirl is invariant under its sign choice. The inclusion \(S\subseteq H\) means that \(H\) contains every ideal stabilizer, while \(\Delta\gamma(H)\) is the purity lost to the additional labels of \(H\).

The input consists of integers \(n\ge1\) and \(0\le t\le n\), a rational promise \(0<a\le1/4\), an accuracy \(\varepsilon>0\), a failure tolerance \(0<\delta<1\), and independent copies of \(\rho\). The promises are \(\nu\le t\) and \(0\le\lambda\le1-\sqrt a\); the actual noise strength is unknown to the learner. The permitted quantum measurements are destructive Bell measurements on pairs of copies. The target is an isotropic \(H\) with \(S\subseteq H\) and \(\Delta\gamma(H)\le\varepsilon\), with success probability at least \(1-\delta\).

The standard Bell measurement measures two copies in the orthonormal basis of maximally entangled vectors
\((I\otimes\sigma_v)|\Phi\rangle\), \(v\in V\), where
\(|\Phi\rangle\coloneqq 2^{-n/2}\sum_x|x,x\rangle\); its raw Bell outcome is the label \(v\).
Each outcome consumes two independent copies of the measured state. Set
\[
 D\coloneqq 2^n,\qquad \alpha\coloneqq (1-\lambda)^2,\qquad r_g\coloneqq \tr(\sigma_g\psi).
\]
Write \(p\) for the raw Bell law on two copies of \(\psi\), and \(U\) for the uniform law on \(V\). The observed law is
\begin{equation}\label{eq:mixture}
 \mathcal D\coloneqq\alpha p+(1-\alpha)U.
\end{equation}
Indeed, expanding \(\rho\otimes\rho\) gives \(\psi\otimes\psi\) weight \(\alpha\), and every other term has a maximally mixed factor, which makes the Bell outcome uniform. We call a label clean when it is drawn from the component \(p\).
Fourier coefficients below are expectations of the characters \(v\mapsto(-1)^{\ip gv}\), \(g\in V\), of the symplectic form.

\subsection{Main results}\label{sec:result-statements}
The following theorem gives polynomial copy complexity throughout \(0\le t\le n\). Its time bound depends exponentially on the supplied upper bound \(t\), while its polynomial factors depend on the fixed noise promise.

\begin{theorem}[Polynomial copies for all stabilizer dimensions]\label{thm:all-noise-polyspace}\label{thm:all-gap}
For every fixed rational \(0<a\le1/4\), there are an algorithm and a polynomial \(p_a\) with the following guarantee. Let \(n\ge1\), \(0\le t\le n\), and let \(\psi\) be a pure \(n\)-qubit state of stabilizer nullity at most \(t\). Suppose that
\[
 \rho=(1-\lambda)\psi+\lambda I/2^n,
 \qquad 0\le\lambda\le1-\sqrt a,
\]
where \(\lambda\) is unknown. Given \(n,t,a\), an accuracy \(\varepsilon>0\), a failure tolerance \(0<\delta<1\), and independent copies of \(\rho\), the algorithm returns, with probability at least \(1-\delta\), an isotropic subspace \(H\le\F_2^{2n}\) satisfying
\[
 S\subseteq H,\qquad \Delta\gamma(H)\le\varepsilon.
\]
On every execution its number of copies of \(\rho\) and its total working storage are at most
\[
 p_a\bigl(n,1/\varepsilon,\log(1/\delta)\bigr),
\]
and its classical bit operations and sequential fair random bits are at most
\[
 2^t p_a\bigl(n,1/\varepsilon,\log(1/\delta)\bigr).
\]
Every quantum measurement is a destructive Bell measurement on two copies. Tolerances are encoded by integer dyadic precision parameters as specified in Section~\ref{mem:model}.
\end{theorem}

Every fixed promised noise ceiling \(\lambda_{\max}<1\) is covered by a rational \(0<a\le\min\{1/4,(1-\lambda_{\max})^2\}\). The theorem includes every \(n\ge1\). Its polynomial degrees and constants, and the dimensions below which it searches all \(4^n\) labels exhaustively, are very large; Section~\ref{sec:conclusion} evaluates these dimensions for the promise \(\lambda\le1/2\).

The purity guarantee yields exact unsigned recovery when the nonstabilizer Pauli expectations are separated from magnitude one.
\begin{corollary}[Exact recovery under a Pauli coefficient gap]\label{cor:exact-recovery}
Under the hypotheses of Theorem~\ref{thm:all-gap}, suppose that a known integer input \(j\ge0\) satisfies
\[
 r_g^2\le1-2^{-j}\qquad\text{for every }g\notin S.
\]
Then \(S\) can be recovered exactly with probability at least \(1-\delta\), using polynomially many copies of \(\rho\) and polynomial working storage in \(n,2^j,\log(1/\delta)\), and at most \(2^t\) times such a polynomial number of classical bit operations and fair bits. The polynomial depends on the fixed noise promise, the bounds hold on every execution, and the quantum measurements are those of Theorem~\ref{thm:all-gap}.
\end{corollary}
\begin{proof}
Run the algorithm of Theorem~\ref{thm:all-gap} with requested accuracy \(\varepsilon\coloneqq 2^{-j}a/4\), using its dyadic rounding convention. Lemma~\ref{lem:purity-consequences} gives, for every \(g\in H\setminus S\),
\[
 \Delta\gamma(H)\ge\frac\alpha2(1-r_g^2)\ge2^{-j-1}a.
\]
This contradicts the successful loss guarantee. Hence \(H=S\), and substituting the accuracy gives the resource bounds.
\end{proof}
When \(t=0\), maximal isotropy already forces \(H=S\) on success, for every requested accuracy. For \(t\ge1\), exact recovery with one copy bound for all admissible states needs a quantitative separation condition (cf.\ \cite[Sections~I and~II.A.1]{TS}). Indeed, \(|00\rangle\) and \(|0\rangle\otimes(\cos\theta|0\rangle+\sin\theta|1\rangle)\), with \(0<\theta<\pi/8\), both have nullity at most one and have different unsigned stabilizer groups. Their \(M\)-copy trace distance is \(\sqrt{1-\cos^{2M}\theta}\le\sqrt M\sin\theta\). Success probability \(1-\delta\) on both inputs would force their output distributions to have total variation at least \(1-2\delta\). For any fixed finite copy cap \(M\) and \(\delta<1/2\), sufficiently small \(\theta\) contradicts contractivity of trace distance under measurement, even for noiseless copies and unrestricted measurements.

Bell outcomes alone cannot determine stabilizer eigenvalue signs. Every computational-basis state has the same raw Bell law: its labels \((0,z)\) are uniform over \(z\in\F_2^n\). Global depolarization preserves this equality by \eqref{eq:mixture}. These states share one unsigned stabilizer group but differ in their stabilizer eigenvalue signs.

\subsection{Proof outline}\label{mem:outline}
The proof separates the construction of candidate isotropic spaces from the final control of purity. A single candidate containing \(S\) suffices: a uniform proposal from an appropriate quotient then finds a new zero-loss direction with probability at least \(2^{-t-1}\) (Proposition~\ref{prop:terminal-list}). A simultaneous verifier built from original Bell outcomes approves such directions and limits the total loss of every accepted sequence. This finder--verifier organization and its purity identity originate in Tang and Steckmann \cite{TS}. Their up-down finder takes the radical of the span of each post-selected batch of Bell difference samples; uniform samples in a batch remove directions from that radical, and the first batch sets the up-down threshold \cite[Section~X]{TS}. The finders below put a label into a span only after filtering or testing it, or when its sampler rarely returns labels outside \(S^\perp\), so the noise enters their guarantees only through the promise \(\alpha\ge a\).

For a fixed positive relative gap \((n-t)/n\), a fixed-degree feature filter constructs a containing isotropic space from polynomially many copies (Section~\ref{sec:linear-gap}). This filter adapts the feature-dependence step of the finite-field mixture algorithm of Chen, De, and Vijayaraghavan \cite[Algorithm~4]{CDV}, and its evaluation-rank input is the rank bound \cite[Lemma~19]{CDV}, whose antecedents are \cite{BHL,KS}. With the verifier and the quotient proposals of \cite{TS}, this range therefore combines known methods; the new work is the polynomial-copy construction when the relative gap tends to zero.

For that range, we split the analysis according to a Fourier moment of one damped Bell-difference distribution; it measures how often two independent sums of samples from this distribution coincide (Section~\ref{sec:moment-framework}). Every state satisfies at least one of the two moment conditions, and the algorithm executes both corresponding procedures. In the low-moment case, a random linear restriction of the Fourier spectrum keeps one missing stabilizer and removes almost all other high-power Fourier mass, and an affine decoder in the style of Blum, Kalai and Wasserman \cite{BKW} recovers that stabilizer from sign biases. One polynomial sample of measurements supports all restriction trials, and a reuse lemma accounts for their dependence. Labels that pass an independent heavy-label test span a subspace inside \(S^\perp\) containing \(S\); its radical is a suitable candidate (Sections~\ref{sec:low-moment} and~\ref{sec:low-span}).

In the high-moment case, two independent sums of samples coincide often, and their common label lies outside \(S^\perp\) with probability at most the inverse of the moment. Translation invariance by \(S\) makes every character nontrivial on \(S\) take the value \(-1\) on that label with probability \(1/2\), which supplies the spread needed for the span to contain \(S\). Bounds on the implemented collision sampler give this spread on typical stored samples of measurements, called caches, and limit the contamination contributed by the remaining ones (Section~\ref{sec:high-moment}). The collision primitive, a low-space search that keeps the working storage polynomial, uses the hash family of Chen, Jin, Williams, and Wu \cite{CJWW} in the version analyzed by Lyu and Zhu \cite{LZ}, together with the collision-finding framework of Beame, Clifford, and Machmouchi \cite{BCM}; the arguments below on caches, fresh random restrictions of the descriptor arrays and deterministic work caps connect that primitive to the actual Bell measurements.

Every procedure returns an isotropic radical on every history. Repetition gives a short list containing at least one suitable radical with high probability, and the final verifier can safely use the entire list. Both small-gap constructions take time \(2^{(1-\eta/2)n}\) times a polynomial, for the constant \(\eta>0\) of Section~\ref{sec:moment-framework}, which depends only on the noise promise; this fits within \(2^t\) because the small-gap range has \(t>(1-\eta/4)n\). Tests on public parameters select an exhaustive fallback in a finite set of dimensions depending only on the fixed noise promise (Section~\ref{sec:assembly}).

\section{Bell identities and observed-state verification}\label{sec:bell-tools}
The raw Bell law exposes the ideal Pauli symmetries, while a verifier built from a separate sample of Bell outcomes controls the purity of the final output. We record the identities and precision conventions needed for both tasks.

\subsection{Bell identities}\label{sec:bell-identities}
The following standard identities expose the structure used below. They use raw Bell outcomes, before taking differences.
\begin{lemma}[Bell identities]\label{lem:bell}
For \(g\ne0\),
\begin{equation}\label{eq:fourier}
 \E_{v\sim\mathcal D}(-1)^{\ip{g}{v}}
   =(-1)^{\pi_Y(g)}\alpha r_g^2.
\end{equation}
Moreover,
\[
 0\le r_g^2\le1,\qquad r_0^2=1,\qquad
 \sum_{g\in V}r_g^2=D.
\]
The law \(p\) is invariant under translation by \(S\) and is supported on
\begin{equation}\label{eq:affine-support}
 A_S\coloneqq \{v:\ip{g}{v}=\pi_Y(g)\text{ for every }g\in S\},
\end{equation}
an affine coset of \(S^\perp\) of dimension \(n+\nu\).
\end{lemma}
\begin{proof}
The Bell Fourier identity is the Pauli-overlap identity of
\cite{HG,TS}; transposition contributes
\(\sigma_g^{\mathsf T}=(-1)^{\pi_Y(g)}\sigma_g\).
Pauli orthogonality gives
\(\sum_g r_g^2=D\tr(\psi^2)=D\).
For \(g\in S\), the pure-state Fourier coefficient has absolute value one, giving the almost-sure equation in \eqref{eq:affine-support}.
The equations are consistent because \(\pi_Y\) is linear on the isotropic space \(S\).
If \(g\notin S^\perp\), conjugation by an anticommuting stabilizer gives \(r_g=0\). Fourier inversion therefore makes \(p\) invariant under every translation in \(S\).
\end{proof}

\subsection{Consequences of the purity guarantee}\label{sec:purity-consequences}
The loss has a direct interpretation in terms of individual approximate symmetries and the disturbance caused by averaging over the output group.
\begin{lemma}[Purity loss and approximate symmetries]\label{lem:purity-consequences}
For every isotropic \(H\),
\begin{equation}\label{eq:twirl-overlap}
 \Delta\gamma(H)=\alpha\left(1-\frac1{|H|}\sum_{h\in H}r_h^2\right).
\end{equation}
Consequently,
\[
 r_g^2\ge1-\frac{2\Delta\gamma(H)}\alpha\quad(g\in H),
 \qquad
 \frac12\|\rho-\mathcal T_H(\rho)\|_1\le\sqrt{\Delta\gamma(H)}.
\]
\end{lemma}
\begin{proof}
The twirl is an orthogonal projection in the Hilbert--Schmidt inner product. Purity of \(\psi\) gives
\[
 \tr\bigl(\mathcal T_H(\psi)^2\bigr)
 =\tr\bigl(\psi\mathcal T_H(\psi)\bigr)
 =\frac1{|H|}\sum_{h\in H}r_h^2.
\]
Global depolarization multiplies the purity loss by \(\alpha\) \cite[proof of Lemma~IV.7]{TS}, proving \eqref{eq:twirl-overlap}. For subspaces \(K\subseteq H\), the range of \(\mathcal T_H\) is contained in that of \(\mathcal T_K\), so \(\Delta\gamma(K)\le\Delta\gamma(H)\) \cite[Observation~IV.1]{TS}. For nonzero \(g\in H\), the two-element twirl therefore yields
\[
 \Delta\gamma(H)\ge\Delta\gamma(\spanf\{g\})
 =\frac\alpha2(1-r_g^2).
\]
The conclusion for \(g=0\) follows from \(r_0=1\). Finally, the trace distance between the pure states \(\psi\) and \(\sigma_h\psi\sigma_h\) is \(\sqrt{1-r_h^2}\). Convexity of the trace norm and Cauchy--Schwarz imply
\[
 \frac12\|\psi-\mathcal T_H(\psi)\|_1
 \le\frac1{|H|}\sum_{h\in H}\sqrt{1-r_h^2}
 \le\sqrt{1-\frac1{|H|}\sum_{h\in H}r_h^2}.
\]
Multiplying by \(1-\lambda=\sqrt\alpha\) proves the observed-state bound.
\end{proof}

Thus a successful output with loss at most \(\varepsilon\) satisfies
\(S\subseteq H\subseteq\{g:r_g^2\ge1-2\varepsilon/a\}\) and causes trace disturbance at most \(\sqrt\varepsilon\). A requested trace-disturbance tolerance \(\tau>0\) can therefore be met by using purity accuracy \(\tau^2\).

\subsection{Precision conventions}\label{mem:model}
The construction below is first given for \(0\le t<n\). The proof in Section~\ref{sec:assembly} then handles \(t=n\) by running it with internal nullity bound \(n-1\); this bound is false only when \(S=\{0\}\), where containment is automatic and the loss and resource bounds hold without it. We express accuracy and confidence through dyadic parameters to make the input encoding and bit-operation bounds explicit.

For the observation model of Section~\ref{sec:learning-model}, we use integer precision inputs. The algorithm receives \(n,t,a\) and positive integers \(e,b\), with
\(\varepsilon\coloneqq 2^{-e}\) and \(\delta\coloneqq 2^{-b}\). We call a quantity public when the algorithm computes it from these inputs alone.
Its bit operations and sequential fair bits, that is, fair coins read once from a one-way source and kept only by copying them into working storage, will be bounded by
\(2^tp_a(n,2^e,b)\), while its original copies and working storage will be a fixed polynomial
in \(n,2^e,b\), for each fixed \(a\).
These bounds hold on every execution, that is, on every sample and random-bit history.
A pair difference is the sum \(v+v'\in V\) of two independent raw labels. Pair differences and all values computed from them are classical data, formed after measurement and stored in polynomial space. We call the measured raw labels and their pair differences original data, and a stored collection of such data, or of values computed from them, a cache.

For arbitrary positive requested tolerances, cap each at \(1/2\), then
choose a dyadic value no larger than that cap and within a factor of two.

Put \(N\coloneqq 2n\).
By Lemma~\ref{lem:bell}, \(p\) is supported on the affine space \(A_S\) of dimension \(n+\nu\le n+t\), and the observed law \(\mathcal D\), like \(p\), is \(S\)-translation invariant.

\subsection{Observed-state verification}\label{mem:verification}
The verifier controls the purity loss accumulated as \(H\) grows. We use one sample of raw Bell outcomes to estimate the loss increments of all candidate extensions with simultaneous error bounds, so that the same data support all later adaptive decisions.

Let \(E_a\) be the least nonnegative integer with \(2^{-E_a}\le a/8\).
Set
\begin{equation}\label{mem:verifier-count}
 E\coloneqq \max(e,E_a),\quad \varepsilon_0\coloneqq 2^{-E},\quad
 h_{\mathrm{step}}\coloneqq \varepsilon_0/n,\quad
 M_{\mathrm{ver}}\coloneqq 8n^2 2^{2E}(4n^2+b+8).
\end{equation}
Draw \(M_{\mathrm{ver}}\) independent raw Bell labels, independently
of finder data. For isotropic \(H\) with basis \(h_1,\ldots,h_k\), use
\[
 Q_H(v)\coloneqq \prod_{j=1}^k\one\{\ip{h_j}v=\pi_Y(h_j)\}.
\]
For nested isotropic \(H\le H'\), the signed estimator \((-1)^{\pi_Y(Z)}(Q_H(Z)-Q_{H'}(Z))\) of \cite[Lemma V.1]{TS} has mean
\begin{equation}\label{mem:purity-estimator}
 \E_{Z\sim\mathcal D}\bigl[(-1)^{\pi_Y(Z)}(Q_H(Z)-Q_{H'}(Z))\bigr]
 =\Delta\gamma(H')-\Delta\gamma(H).
\end{equation}
The factor \((-1)^{\pi_Y(Z)}\) is the swap eigenvalue of the Bell state labeled by \(Z\), as the proof below shows.
Here \(\Delta\gamma\) is the additive purity loss of the observed state \(\rho\), and \(\mathcal T_H\) is the twirl of Section~\ref{sec:learning-model}, the Hilbert--Schmidt projection onto the span of \(\{\sigma_g:g\in H^\perp\}\).

For completeness we include a short proof of the identity \eqref{mem:purity-estimator} of \cite[Lemma~V.1]{TS}, which is proved there by a different argument.
Since \(\pi_Y\) is linear on the isotropic space \(H\),
\(Q_H(v)=|H|^{-1}\sum_{h\in H}(-1)^{\ip hv+\pi_Y(h)}\).
The Bell state labeled by \(v\) is an eigenvector of the swap \(\mathrm{SWAP}\) of the two copies with eigenvalue
\((-1)^{\pi_Y(v)}\), and of \(\sigma_h\otimes\sigma_h^{\mathsf T}\) with eigenvalue \((-1)^{\ip hv}\).
Using \(\sigma_h^{\mathsf T}=(-1)^{\pi_Y(h)}\sigma_h\) and \(\tr(\mathrm{SWAP}\,(A\otimes B))=\tr(AB)\),
the signs \((-1)^{\pi_Y(h)}\) cancel and
\[
 \E\bigl[(-1)^{\pi_Y(Z)}Q_H(Z)\bigr]
 =\frac1{|H|}\sum_{h\in H}\tr(\sigma_h\rho\sigma_h\rho)
 =\tr\bigl(\mathcal T_H(\rho)\rho\bigr)
 =\tr\bigl(\mathcal T_H(\rho)^2\bigr),
\]
the last step because \(\mathcal T_H\) is an orthogonal projection.
Subtracting the same identity for \(H'\) gives \eqref{mem:purity-estimator}.

Tang and Steckmann's uniform verification \cite[Lemma~V.3 and Theorem~V.4]{TS} gives a sharper sample size; we use the following elementary union bound to keep the resource accounting explicit. There are at most \(2^{4n^2}\) ordered isotropic pairs, since each isotropic space is spanned by some \(n\) vectors of \(V\). Hoeffding gives
an event \(\mathcal E_{\mathrm{ver}}\) on which every empirical
increment is accurate within \(h_{\mathrm{step}}/2\), with
\begin{align}
 \Prob(\mathcal E_{\mathrm{ver}}^c)
 &\le2^{4n^2+1}\exp(-M_{\mathrm{ver}}h_{\mathrm{step}}^2/8)\notag\\
 &=2^{4n^2+1}\exp(-(4n^2+b+8))\le\delta/128.
 \label{mem:verifier-failure}
\end{align}
Initialize \(H\coloneqq 0\). Call \(g\) eligible when \(g\in H^\perp\setminus H\). Approve an eligible \(g\), replacing \(H\) by \(H+\spanf\{g\}\), exactly when its empirical incremental loss is at most \(h_{\mathrm{step}}/2\). Since \(g\) is orthogonal to \(H\) and to itself, every approval keeps \(H\) isotropic, on every history. On \(\mathcal E_{\mathrm{ver}}\) each approval costs at most \(h_{\mathrm{step}}\), and each eligible \(g\) of zero incremental loss is approved; we call such \(g\) a zero-loss extension. At most \(n\) approvals
give total loss at most \(\varepsilon_0\). The verifier only judges proposed labels; Proposition~\ref{prop:terminal-list} and the exhaustive scans of Sections~\ref{sec:linear-gap} and~\ref{sec:finite-fallback} generate them and decide when to stop.

Since \(\varepsilon_0\le a/8<\alpha/4\), the compatibility lemma \cite[Lemma IV.7]{TS} gives
\(H\subseteq S^\perp\) throughout. For completeness we include the short argument.
Global depolarization multiplies the pure-state loss by \(\alpha\), and
\(\psi=D^{-1}\sum_g r_g\sigma_g\) gives the pure-state loss
\(1-\tr(\mathcal T_H(\psi)^2)=D^{-1}\sum_{g\notin H^\perp}r_g^2\).
The labels of \(S^\perp\) carry all of \(\sum_g r_g^2=D\), by the proof of Lemma~\ref{lem:bell}.
If some \(h\in H\) has \(\ip h{s_0}=1\) for some \(s_0\in S\), the involution
\(g\mapsto g+s_0\) of \(S^\perp\) satisfies \(r_{g+s_0}^2=r_g^2\), because \(\sigma_{s_0}\psi=\pm\psi\),
and \(\ip h{g+s_0}\ne\ip hg\).
At least half of this mass therefore lies outside \(H^\perp\), so
\(\Delta\gamma(H)\ge\alpha/2>\varepsilon_0\), which is impossible for an accepted \(H\) on
\(\mathcal E_{\mathrm{ver}}\). Thus every \(g\in S\setminus H\)
and every representative of a nonzero coset in \((S+H)/H\) is eligible
and has zero incremental loss. If \(\dim H=n\), then
\(S\subseteq H^\perp=H\). This one simultaneous event covers all later
adaptive queries.

\subsection{A necessary dependence on the noise level}\label{sec:noise-lower-bound}
Even the one-qubit task requires increasingly many Bell outcomes as the observed state approaches complete depolarization. The following two-state calculation gives a quantitative restriction on any bound uniform in the noise level.

\begin{proposition}[Raw Bell sample lower bound]\label{prop:bell-noise-lower-bound}
Let \(0<\alpha<1\) and \(0<\delta<1/2\). Suppose a learner receives only outcomes of independent destructive standard Bell measurements on disjoint pairs of copies, with at most \(M\) outcomes on every execution, and, with probability at least \(1-\delta\), outputs an isotropic subspace containing the unsigned stabilizer group of the underlying pure state, for both global depolarizations of \(|0\rangle\) and \(|+\rangle\),
\[
 \rho_Z\coloneqq\frac{I+\sqrt\alpha Z}{2},\qquad
 \rho_X\coloneqq\frac{I+\sqrt\alpha X}{2}.
\]
Put \(d_\delta\coloneqq(1-2\delta)\ln((1-\delta)/\delta)\). Then
\begin{equation}\label{eq:bell-noise-lower-bound}
 M\ge
 \frac{d_\delta}{(\alpha/2)\ln((1+\alpha)/(1-\alpha))}
 \ge\frac{1-\alpha^2}{\alpha^2}\,d_\delta.
\end{equation}
In particular, if \(\alpha\le1/2\) and \(\delta\le1/4\), the worst-case original-copy budget satisfies
\[
 2M\ge\frac38(1-\lambda)^{-4}\ln(1/\delta),
 \qquad \lambda=1-\sqrt\alpha.
\]
\end{proposition}
\begin{proof}
These are the global depolarizations of \(|0\rangle\) and \(|+\rangle\), with \(n=1,t=0\). Their unsigned stabilizer spaces are the distinct maximal isotropic lines \(\spanf\{Z\}\) and \(\spanf\{X\}\). Thus successful containment forces these respective outputs. Both have zero purity loss on the corresponding state.

In Bell-state order \((\Phi^+,\Phi^-,\Psi^+,\Psi^-)\), direct Born probabilities are
\[
 P_Z\coloneqq\tfrac14(1+\alpha,1+\alpha,1-\alpha,1-\alpha),\qquad
 P_X\coloneqq\tfrac14(1+\alpha,1-\alpha,1+\alpha,1-\alpha).
\]
Only the second and third terms contribute to their relative entropy, giving
\[
 D_{\mathrm{KL}}(P_Z\Vert P_X)
 =\frac\alpha2\ln\frac{1+\alpha}{1-\alpha}.
\]
Pad an early-stopping execution with ignored outcomes to length \(M\). The output, including arbitrary independent classical randomness, is a randomized function of this independent sample. Let \(A\) be the event that the output is \(\spanf\{Z\}\). Its probabilities satisfy \(p\coloneqq P_Z^M(A)\ge1-\delta\) and \(q\coloneqq P_X^M(A)\le\delta\), where the common randomized decision is included. The log-sum inequality applied to this decision and its complement gives
\[
 M D_{\mathrm{KL}}(P_Z\Vert P_X)
 \ge p\ln\frac pq+(1-p)\ln\frac{1-p}{1-q}
 \ge d_\delta.
\]
The last expression is minimized over \(p\ge1-\delta\), \(q\le\delta\) at \(p=1-\delta,q=\delta\): its derivative in \(p\) is positive for \(p>q\), and its derivative in \(q\) is negative. Finally,
\[
 \frac12\ln\frac{1+\alpha}{1-\alpha}
 =\int_0^\alpha\frac{du}{1-u^2}
 \le\frac\alpha{1-\alpha^2},
\]
which proves \eqref{eq:bell-noise-lower-bound}. For the last assertion use \(1-\alpha^2\ge3/4\), \(d_\delta\ge\tfrac14\ln(1/\delta)\) when \(\delta\le1/4\), and two original copies per Bell outcome.
\end{proof}
This dependence on noise is compatible with a polynomial bound in \(1/(1-\lambda)\); the present upper bound fixes a positive noise promise before choosing its polynomial.

\subsection{Verified proposals from an arbitrary list}\label{sec:terminal-list}
The following elementary construction extends the quotient proposal rule of Tang and Steckmann \cite[Algorithm~7 and Lemma~VII.5]{TS} to a list of isotropic spaces. The list may also contain arbitrary unsuccessful outputs.

\begin{proposition}[Terminal list procedure]\label{prop:terminal-list}
Let \(R_1,\ldots,R_q\) be isotropic subspaces. There is a procedure that always returns an isotropic \(H\), and on the simultaneous verifier event \(\mathcal E_{\mathrm{ver}}\) of Section~\ref{mem:verification} satisfies \(\Delta\gamma(H)\le\varepsilon_0\). Under the standing promise \(\nu\le t\), if at least one list member contains \(S\), then, conditional on this event, \(H\) fails to contain \(S\) with probability at most \(\delta/64\). It uses the existing verifier cache, polynomial additional storage in \(n,q,b\), and at most
\[
 (n+1)q\,2^{t+1}\bigl(b+\lceil\log_2(n+1)\rceil+6\bigr)
\]
proposals on every history.
\end{proposition}
\begin{proof}
Start \(H\coloneqq0\). At each stage and for each list member, compute
\[
 J_i\coloneqq H+(R_i\cap H^\perp),\qquad
 K_{\mathrm{term}}\coloneqq b+\lceil\log_2(n+1)\rceil+6.
\]
The sum \(J_i\) is isotropic: its two summands are isotropic and commute with one another. Draw at most \(2^{t+1}K_{\mathrm{term}}\) uniform cosets of \(J_i/H\), using fair coefficients in a computed quotient basis. Count zero cosets as trials and skip them. Every other representative is eligible for the verifier. Restart the stage after an approval. Return \(H\) when a stage exhausts all trials, or when \(\dim H=n\).

On \(\mathcal E_{\mathrm{ver}}\), each approval has true incremental loss at most \(\varepsilon_0/n\), and every zero-loss eligible direction is approved. The simultaneous event covers the adaptive sequence, which has at most \(n\) approvals. It follows that \(\Delta\gamma(H)\le\varepsilon_0\) and \(H\subseteq S^\perp\).

Fix a stage at which \(S\nsubseteq H\), and a good list member \(R_i\supseteq S\). Compatibility gives \(S+H\subseteq J_i\). Put
\[
 u\coloneqq\dim((S+H)/H)\ge1,\qquad v\coloneqq\dim(J_i/(S+H)).
\]
Since \(\dim J_i\le n\) and \(\dim S\ge n-t\), we have \(v\le t\). The fraction of uniform quotient cosets representing a zero-loss extension in \((S+H)/H\) is
\[
 \frac{2^u-1}{2^{u+v}}\ge2^{-t-1}.
\]
The prescribed trials therefore fail to make progress with probability at most \(\exp(-K_{\mathrm{term}})\). An earlier approval from another member is already safe progress. The union over at most \(n+1\) stages is at most \(\delta/64\). If rank \(n\) is reached, maximal isotropy and compatibility give \(S\subseteq H^\perp=H\). Binary elimination, uniform quotient sampling and each empirical test require polynomial work and workspace per proposal, proving the claimed caps.
\end{proof}

\section{Filtering at a fixed relative dimension gap}\label{sec:linear-gap}\label{mem:fixed-filter}
At a fixed positive relative dimension gap, a fixed-degree feature filter
produces isotropic spaces, one of which contains \(S\) with high probability.
Uniform quotient proposals from these spaces then yield the required
verified output. We prove this case as a complete learner, including
small dimensions, where it searches exhaustively, so that
Section~\ref{sec:assembly} can use it as a separate branch. Its proof uses
the terminal list procedure of Proposition~\ref{prop:terminal-list}, whose
proof uses only the verifier of Section~\ref{mem:verification}.

\begin{proposition}[Every fixed relative gap]\label{prop:linear-gap}\label{cor:linear-gap}\label{mem:prop-fixed-gap}
Fix rational numbers \(0<a\le1/4\) and \(0<c<1\), and put
\(a_c\coloneqq\min\{a,c\}/2\).
There is a learner, depending only on \(a,c\), which for every
\(n\ge1\) and \(0\le t<n\) with \(n-t\ge cn\), in the observation model
with \(\alpha\ge a\), returns with probability at least \(1-\delta\) an
isotropic \(H\supseteq S\) with \(\Delta\gamma(H)\le\varepsilon\).
For each fixed \(a,c\), its original copies and total working storage are
bounded on every execution by a fixed polynomial in
\(n,1/\varepsilon,\log(1/\delta)\), and its bit operations and fair random
bits by \(2^t\) times such a polynomial. It uses destructive two-copy
Bell measurements and the reduced noise promise \(a_c\).
\end{proposition}

Use the dyadic precision convention of Section~\ref{mem:model} and the
verifier of Section~\ref{mem:verification}, with \(a_c\) in place of \(a\).
Thus \(\varepsilon_0\le a_c/8\), and \(M_{\mathrm{ver}}\) denotes its
sample count with this substitution. Define the fixed constants
\begin{equation}\label{mem:fixed-constants}
 \mu\coloneqq a_c^2/16,\qquad
 k_f\coloneqq\lceil\log_2(16/a_c)\rceil,\qquad
 \ell_f\coloneqq\lceil4k_f/\mu\rceil.
\end{equation}
Let \(N_f\) be the least power of two satisfying
\begin{align}
 N_f&\ge\max\{4\ell_f/\mu,\,8\ell_f^2/a_c,\,100/a_c\},
 \label{mem:filter-cutoff-size}\\
 \ell_f\lceil\log_2(N_f+1)\rceil
       -\frac{a_cN_f}{2\ell_f}+\ell_f+1&\le-7.
 \label{mem:filter-cutoff-tail}
\end{align}
A doubling search terminates because the negative linear term dominates
the logarithm. All ceilings and comparisons are exact integer or rational
operations, and these constants depend only on \(a,c\).
Since \(a_c\le c/2\) and \(c<1\), we have
\(\mu\le c^2/64<c\). Consequently the gap
\(n-t\ge cn\) satisfies \(n-t>\mu n\). This section uses \(\mu\) only through this inequality, \(\ell_f\ge4k_f/\mu\) and \(N_f\ge4\ell_f/\mu\), so any fixed \(0<\mu<c\) would serve; these constants are sufficient.

For now assume \(N=2n\ge N_f\).
By \eqref{eq:mixture}, every raw Bell label is independently clean with
probability \(\alpha\ge a_c\), in which case it has law \(p\), and
otherwise has the uniform law \(U\). The latent clean indices in the
proof refer to this decomposition.

Write
\[
 B(j,\ell)\coloneqq \sum_{i=0}^{\min(j,\ell)}\binom ji,
\]
and let the feature vector \(\Phi_\ell(x)\) list all squarefree monomials of degree at most
\(\ell\), including the constant.
Thus \(\Phi_\ell:\F_2^N\to\F_2^{B(N,\ell)}\).

\label{mem:feature-rank-input}

We use the following evaluation-rank fact \cite[Lemma~19]{CDV}: a set of at least \(2^z\) distinct binary points has feature-span dimension at least \(B(z,\ell)\), for integers \(0\le z\le N\). Chen, De, and Vijayaraghavan derive this from the polynomial-rank lemma of Ben-Eliezer, Hod, and Lovett \cite{BHL}; see \cite[Lemma~18]{CDV}. An equivalent earlier inclusion-matrix rank bound appears in \cite[Theorem~1.5]{KS}. The filter of Lemma~\ref{mem:lem-radical} is the feature-dependence step of \cite[Algorithm~4]{CDV}, and its proof adapts \cite[proof of Theorem~2]{CDV} to an affine clean component, uniform contamination and explicit constants.

\begin{lemma}[Radicals at a fixed relative gap]\label{mem:lem-radical}
Assume \(N\ge N_f\) and \(n-t>\mu n\). Draw \(B(N,\ell_f)\) raw Bell labels, retain exactly the indices whose feature columns participate in a nonzero linear dependence among all these columns, and return the radical of the span of all pairwise differences of the retained labels, or zero if no label is retained. Every run returns an isotropic subspace, whose dimension is therefore at most \(n\le\dim S+t\). With probability at least \(61/64\), the returned subspace contains \(S\). One run uses \(B(N,\ell_f)\) labels, and its bit operations and stored bits are polynomial in \(n\) for fixed \(a,c\).
\end{lemma}
\begin{proof}
The letters \(F\), \(R\), \(G\) and \(f\) below are local to this proof. Use
\[
 \ell\coloneqq \ell_f,\quad F\coloneqq B(N,\ell),\quad d\coloneqq n+t,\quad
 R\coloneqq B(d,\ell),\quad
 z\coloneqq \left\lceil N-\frac{a_cN}{2\ell}+\ell\right\rceil.
\]
The gap \(N-d=n-t>\mu n\) and the cutoff give
\(\ell\le(N-d)/2\) and \(\ell\le z\le N\).
Clean labels lie in the affine space \(A_S\), of dimension at most \(d\); substituting an affine parametrization of \(A_S\) turns each feature into a polynomial of degree at most \(\ell\) in at most \(d\) variables, so the clean feature columns span at most \(R\) dimensions, and
\begin{equation}\label{mem:filter-rank}
 \frac RF\le\left(\frac{d+\ell}{N}\right)^\ell
 \le(1-\mu/4)^\ell
 \le e^{-\ell\mu/4}\le e^{-k_f}\le2^{-k_f}\le a_c/16.
\end{equation}
The first inequality holds because \(R\le\binom{d+\ell}{\ell}\) by Vandermonde's identity, \(F\ge\binom N\ell\), and \(\binom{d+\ell}{\ell}/\binom N\ell=\prod_{j=0}^{\ell-1}(d+\ell-j)/(N-j)\), where each factor is at most \((d+\ell)/N\) because \(\ell\le(N-d)/2\) gives \(d+\ell\le N\).

The feature map \(\Phi_\ell\) separates the two components of the \(F\) drawn labels: the clean columns, about \(\alpha F\ge a_cF\) of the \(F\) columns, span at most \(R\) dimensions, and the degree makes \(R\le a_cF/16\) by \eqref{mem:filter-rank}, so most clean columns lie in dependences; a contamination column joins a dependence only for labels in a set of fewer than \(2^z\) points, as shown next.
If \(G\) is the latent clean count, then
\[
 \Prob(G<7a_cF/8)\le e^{-a_c^2F/32}.
\]
On the complementary count event, the span of the features \(\Phi_\ell(x)\), \(x\in A_S\),
and all but one contamination column has rank at most
\[
 R+F-G-1<(1-13a_c/16)F<(1-a_c/2)F.
\]
But \(\binom zi\ge((z-\ell)/N)^\ell\binom Ni\) for \(0\le i\le\ell\), since \((z-j)/(N-j)\ge(z-\ell)/N\) for \(0\le j<\ell\), and summing over \(i\) gives
\[
 \frac{B(z,\ell)}F
 \ge\left(\frac{z-\ell}{N}\right)^\ell
 \ge\left(1-\frac{a_c}{2\ell}\right)^\ell\ge1-a_c/2.
\]
By the evaluation-rank fact, the feature preimage of the span formed from the clean support and the other contamination columns therefore has fewer than \(2^z\) points.
Given the latent indices, each contamination label is uniform and independent of the other labels, so a union bound over contamination indices gives failure at most \(F2^{z-N}\).
Outside this event no contamination column is retained.

At most \(R\) clean columns are deleted, because deleted columns lie in no nonzero dependence, so the deleted clean columns are linearly independent in the clean feature span, of dimension at most \(R\). Hence a fraction at least \(1-R/G\ge13/14>9/10\) of the clean indices
remains. Every affine hyperplane whose defining linear functional is nonzero on \(S\) has clean mass \(1/2\), by the \(S\)-translation invariance of \(p\) (Lemma~\ref{lem:bell}). Conditional on the latent clean indices, Hoeffding and a union bound show that the event \(\mathcal E_{\mathrm{span}}\), on which no such hyperplane contains at least \(9G/10\) clean labels, fails with probability at most
\[
 2^{N+1}e^{-8G/25}\le2^{N+1}e^{-7a_cF/25}.
\]
Outside this failure event and the earlier failures, more than \(9G/10\) clean labels are retained. If the span \(V_0\) of their pairwise differences did not contain \(S\), a linear functional vanishing on \(V_0\) but nonzero on \(S\) would give a forbidden hyperplane containing them all. Thus \(S\subseteq V_0\). The retained labels lie in \(A_S\), so \(V_0\subseteq S^\perp\), equivalently \(S\subseteq V_0^\perp\). Hence \(\rad(V_0)=V_0\cap V_0^\perp\) contains \(S\) and, being isotropic, has dimension at most \(n\le\dim S+t\).
On every run, including failed ones, the returned subspace, a radical or zero, is isotropic.

Each failure term is at most \(1/64\) for \(N\ge N_f\).
Indeed \(F\ge N^2/2\), \(N\ge100/a_c\), and
\(a_c^2F/32\ge10000/64>6\).
For the contamination term \(F2^{z-N}\), put
\[
 f(N)\coloneqq \ell\log_2(N+1)-a_cN/(2\ell)+\ell+1.
\]
Condition \eqref{mem:filter-cutoff-tail} gives \(f(N_f)\le-7\).
For \(N\ge N_f\), its derivative is less than
\(2\ell/(N+1)-a_c/(2\ell)<0\), using \(N_f\ge8\ell^2/a_c\).
Since \(F\le(N+1)^\ell\), we obtain \(F2^{z-N}\le2^{-7}\).
Finally \(7a_cF/25\ge7a_cN^2/50\ge14N\ge N+7\) bounds the hyperplane term.
One run succeeds with probability at least \(61/64\).

Forming the feature matrix and computing its right kernel once uses
\[
 F\text{ observations},\qquad
 O(F^3+\ell F^2+N^2F+N^3)\text{ binary operations},\qquad
 O(F^2+NF)\text{ bits}.
\]
Here \(\ell\) depends only on fixed \(a,c\), so every quantity is polynomial
in \(n\).
\end{proof}

\begin{proof}[Proof of Proposition~\ref{prop:linear-gap}]
First suppose \(N\ge N_f\). Generate
\(K_{\mathrm{pre}}\coloneqq b+6\) independent radicals by
Lemma~\ref{mem:lem-radical}, independently of the verifier data, and keep
all their bases. The probability that none contains \(S\) is at most
\[
 (3/64)^{K_{\mathrm{pre}}}\le2^{-K_{\mathrm{pre}}}=\delta/64.
\]
Apply the terminal list procedure of Proposition~\ref{prop:terminal-list} to this
list, with
\begin{equation}\label{mem:terminal-counts}
 K_{\mathrm{term}}\coloneqq b+\lceil\log_2(n+1)\rceil+6,\qquad
 L_{\mathrm{term}}\coloneqq2^{t+1}K_{\mathrm{term}}.
\end{equation}
Every list member is isotropic on every execution. On the event that
one contains \(S\), Proposition~\ref{prop:terminal-list} applies with the same
supplied upper bound \(t\). Its failure budget, including the independent
purity verifier, is at most \(\delta/128+\delta/64\).
Thus total failure is at most
\(\delta/128+\delta/64+\delta/64<\delta\).
The raw Bell count is
\(M_{\mathrm{ver}}+K_{\mathrm{pre}}B(N,\ell_f)\), and the original copy
count is twice this quantity. The terminal list procedure uses at most
\((n+1)K_{\mathrm{pre}}L_{\mathrm{term}}\) candidate trials.

If \(N<N_f\), acquire only the verifier sample and start at \(H\coloneqq0\).
Enumerate all \(4^n\) labels, approving only eligible extensions under
the same verifier rule. Restart the scan after an approval; stop after
a scan with no approval or when \(\dim H=n\). On
\(\mathcal E_{\mathrm{ver}}\), each missing stabilizer is an eligible
zero-loss extension, so termination implies \(S\subseteq H\). There
are at most \(n+1\) scans on every history. The copy count is
\(2M_{\mathrm{ver}}\), and the \(4^n\) work factor is bounded by
\(2^{N_f}\), a constant depending only on \(a,c\).

In either case the verifier gives
\(\Delta\gamma(H)\le\varepsilon_0\le\varepsilon\) on its simultaneous
event. Eligibility preserves isotropy even when that event fails.
Since \(\ell_f\) is fixed, \(B(N,\ell_f)\le(N+1)^{\ell_f}\) is
polynomial. Filters are run sequentially, storing only their radical
bases after each run; the feature matrix, current basis computations,
verifier data and counters have polynomial storage. Quotient proposals
are generated and tested one at a time with at most \(n\) fair bits
per trial. Their explicit caps and the finite scan bound give
\(2^t\) times polynomial bit work and sequential fair bits on every
execution. Dyadic rounding gives the stated guarantees for arbitrary
requested tolerances.
\end{proof}

\section{A common moment for the small-gap constructions}\label{sec:moment-framework}
The two constructions in the following sections use the same distribution and the same Fourier moment. One isolates a stabilizer frequency when the moment is small; the other uses label collisions when it is large. Their analysis ranges overlap, so the learner can run both and retain all candidate radicals.

\subsection{The working law}
For \(g\in V\), put \(B_g\coloneqq r_g^2\) and \(G_g\coloneqq B_g^2\). Write \(P\coloneqq p*p\) for the Bell difference sampling distribution of \(\psi\), the law of the sum of two independent samples from \(p\). Bell difference sampling and its fourth-moment representation are established primitives \cite{GNW,GIKL24}. Convolution and Fourier transforms are on \(V\), with characters \((-1)^{\ip gx}\) and Fourier coefficients normalized as expectations. Lemma~\ref{lem:bell} gives
\[
 \widehat P(g)=G_g,\qquad 0\le B_g\le1,\qquad
 \sum_g B_g=2^n.
\]
The law \(P\) is supported on \(S^\perp\) and invariant under translation by \(S\). Indeed, the sum of two elements of the affine support \(A_S\) lies in \(S^\perp\), and translation invariance is preserved by convolution. Also \(B_{g+s}=B_g\) for \(s\in S\), since multiplying a Pauli by an ideal stabilizer preserves the absolute value of its expectation.

The sum of two independent observed raw Bell labels costs four original copies and has law
\[
 \alpha^2P+(1-\alpha^2)U.
\]
With probability \(1/2\), replace such a label by an independent uniform label. The resulting working law is
\begin{equation}\label{eq:working-pair-law}
 R\coloneqq\beta P+(1-\beta)U,\qquad
 \beta\coloneqq\alpha^2/2,\qquad \beta_{\min}\coloneqq a^2/2\le\beta\le1/2.
\end{equation}
The half-uniform replacement gives the working law full support, with \(R(v)\ge1/(2|V|)\) for every \(v\in V\); in particular, the conditional sign mean \eqref{aff:posterior} of Appendix~\ref{sec:affine-cache} is then defined at every input. Let \(f\coloneqq\widehat R\). Thus \(f(0)=1\) and \(f(g)=\beta G_g\) for \(g\ne0\). The working law is \(S\)-translation invariant. All later estimates hold throughout \(\beta_{\min}\le\beta\le1\). The heavy-label test and the verifier use separate original raw Bell outcomes from \(\rho\); damping is applied only to working data.

\subsection{Public parameters and overlapping promises}
The constants \(g_0,m_0,L,\eta,\kappa,P_0\) below depend only on the fixed noise promise, while \(\ell_n\) and \(w\) also depend on \(n\). Define
\begin{equation}\label{eq:moment-parameters}
\begin{aligned}
 g_0&\coloneqq\lceil\log_2(1/\beta_{\min})\rceil,&
 m_0&\coloneqq8g_0+8,& L&\coloneqq2^{m_0}-1,\\
 \eta&\coloneqq\frac1{100L},& \kappa&\coloneqq\frac\eta4,&
 P_0&\coloneqq\left\lceil\frac{1024(g_0+1)}\eta\right\rceil,\\
 \ell_n&\coloneqq\lceil\log_2(n+2)\rceil,&
 w&\coloneqq\left\lfloor\frac{n}{P_0\ell_n}\right\rfloor.&
\end{aligned}
\end{equation}
Here \(w\) is the number of working labels in each grouped sum, \(\eta\) sets the moment thresholds below, \(\kappa\) is the relative gap below which the small-gap constructions are used, \(L=2^{m_0}-1\) is the length of the Hamming code in the low-moment decoder, and \(2^{g_0}\ge1/\beta_{\min}\); \(m_0\) and \(P_0\) are chosen so that the low-moment cost exponent stays below \(n-\lceil n/L\rceil\) (Lemma~\ref{low:eventual}). For \(w\ge1\), define the analysis quantity and a public lower threshold
\begin{equation}\label{eq:moment-split}
 A\coloneqq\sum_{g\in V}f(g)^{2w},\qquad A_{\min}\coloneqq2^{\lfloor\eta n\rfloor}.
\end{equation}
Every state satisfies at least one of the following promises:
\[
 \text{low moment: }A\le2^{\eta n},\qquad
 \text{high moment: }A\ge A_{\min}.
\]
Both promises are conditions on the same quantity \(A\), which the state determines through the working law: by Parseval, \(A/|V|\) is the probability that two independent sums of \(w\) working labels coincide. The algorithm uses the displayed public parameters and executes both constructions; the unknown value of \(A\) enters only their success analysis.

The supplied bound \(t\) determines the first branch: for \(n-t\ge\kappa n\), use Proposition~\ref{prop:linear-gap} with \(c=\kappa\). For \(n-t<\kappa n\), use the constructions of Sections~\ref{sec:low-moment} and~\ref{sec:high-moment}. Each has public tests stated in its proof. At fixed \(a\), all of those tests hold for every sufficiently large \(n\). When a test fails, the finite-dimensional fallback of Section~\ref{sec:finite-fallback} applies. In particular, the algorithm tests \(w\ge1\) before using a formula involving \(1/w\).

The collision procedures use the constant \(a_0=1/61440\) of Lemma~\ref{coll:explicit-lower}: in the collision walk analyzed by Lyu and Zhu \cite{LZ} on an array with \(F\) ordered index pairs of equal value, including equal indices, both indices of each collision are visited with probability at least \(a_0/F\). They also use a computable bound \(Q\) on the work per walk step, polynomial in the logarithm of the array length and the cost of evaluating one array entry (Appendix~\ref{sec:collision}). The remaining work coefficients, polynomial degrees and finite cutoffs depend on the fixed noise promise.

\section{Isolating a stabilizer when the moment is small}\label{sec:low-moment}

Suppose that
\begin{equation}\label{low:promise}
 A=\sum_{g\in V}f(g)^{2w}\le 2^{\eta n}.
\end{equation}
When the public tests below hold, for any fixed space \(C\subseteq S^\perp\) with \(S\nsubseteq C\), we construct a procedure
whose candidate labels include an element of \(S\setminus C\) with probability greater than \(7/8\). Section~\ref{sec:low-span}
iterates it, with an independent heavy-label test, to build such a space
that contains \(S\); on the simultaneous good event for that test, every passing label has \(B_g>0\) and hence lies in \(S^\perp\) by the proof of Lemma~\ref{lem:bell}, so the iteration keeps \(C\subseteq S^\perp\). A random frequency
restriction preserves one missing stabilizer and removes almost all of the
remaining high-power Fourier mass. On that event, the affine decoder, a collision search with Hamming-coded keys,
recovers the stabilizer from affine equations. The restriction succeeds
rarely, so the procedure repeats it on one saved cache.
The last part of the proof justifies that reuse.

Throughout this section, \(G_g=B_g^2=\widehat P(g)\), and
\(\chi_g(y)\coloneqq(-1)^{\ip gy}\). The working law is
\(R=\beta P+(1-\beta)U\), with
\(\beta_{\min}=a^2/2\le\beta\le1/2\), as in the common setup.

\subsection{A restriction and a coordinate chart}

Fix for the analysis a label \(s\in S\setminus C\). Put
\begin{equation}\label{low:filter-size}
 a_n\coloneqq\lceil\eta n\rceil,\qquad
 r\coloneqq\min\{j\ge0:2^j\ge64\beta_{\min}^{-2w}2^{a_n}\}.
\end{equation}
Draw an independent uniform binary \(r\)-by-\(2n\) matrix \(M\), let
\(F\coloneqq\ker M\), and let \(W\coloneqq F^\perp\).
Add an independent uniform \(W\)-vector to every working row. For fixed \(M\)
the resulting population laws are
\begin{equation}\label{low:masked-law}
 P_F\coloneqq P*U_W,\qquad R_F\coloneqq\beta P_F+(1-\beta)U,\qquad
 \widehat P_F(g)=G_g\one_{g\in F}.
\end{equation}
A basis of \(W\) is computed using the symplectic form; independent fair
basis coefficients generate each mask. This construction allows deficient
rank of \(M\).

\begin{lemma}\label{low:isolation}
Under \eqref{low:promise}, with probability at least
\((7/8)2^{-r}\), the restriction satisfies
\begin{equation}\label{low:isolated}
 s\in F,\qquad
 \sum_{g\in F\setminus\{0,s\}}G_g^{2w}\le\frac18.
\end{equation}
Independently draw \(z_0\) uniformly from \(C^\perp\), including zero.
With probability exactly \(1/2\), \(\ip{s}{z_0}=1\). When \(z_0\ne0\), choose
a frequency basis \(\phi_1,\ldots,\phi_k\) of
\(\mathcal L\coloneqq z_0^\perp\), where \(k\coloneqq2n-1\), and choose \(v_0\) with
\(\ip{v_0}{z_0}=1\). On the event \(\ip{s}{z_0}=1\), there is a unique \(u\in\F_2^k\) with
\(s=v_0+\sum_j u_j\phi_j\), and every label of that affine form is outside \(C\).

If both events hold, then for a row \(Y\in V\) write \(X\coloneqq(\ip{\phi_j}{Y})_{j=1}^k\) for its coordinates, and \((P_F)_X,(R_F)_X\) for the laws of \(X\) when \(Y\) has law \(P_F\) or \(R_F\). Put
\(D_X\coloneqq2^k\), \(p_c\coloneqq(P_F)_X^{*c}\), and \(r_c\coloneqq(R_F)_X^{*c}\).
For every \(c\ge2w\) and every \(x\in\F_2^k\),
\begin{align}\label{low:flat-density}
 \frac{7}{8D_X}\le p_c(x)\le\frac9{8D_X},\qquad
 \frac{1-\beta^c/8}{D_X}\le r_c(x)\le
 \frac{1+\beta^c/8}{D_X},\\
 \gamma_c(x)\coloneqq\frac{\beta^cp_c(x)}{r_c(x)}\ge\frac{\beta^c}{2},
 \qquad
 \max\left\{\frac{r_c(0)}{r_c(x)},\frac{p_c(0)}{p_c(x)}\right\}\le2.
 \nonumber
\end{align}
Moreover \(\sum_{g\in\mathcal L}\widehat R_F(g)^{2w}\le9/8\).
\end{lemma}

\begin{proof}
For \(g\notin\{0,s\}\), independence of two binary linear functionals gives
\(\Prob[Ms=0]=2^{-r}\) and
\(\Prob[Mg=0\mid Ms=0]=2^{-r}\). Also
\(\sum_{g\ne0,s}G_g^{2w}\le\beta_{\min}^{-2w}2^{a_n}\).
Conditional on \(Ms=0\), the expected sum in \eqref{low:isolated} is at most
\(1/64\), so Markov's inequality proves the first assertion.
The functional \(\ip{s}{\cdot}\) is nonzero on \(C^\perp\), since
\(s\notin C\), proving the probability \(1/2\).
Since \(C\subseteq z_0^\perp\), all affine candidates lie outside \(C\).

The masked clean law is supported on \(s^\perp\), and its Fourier
coefficients satisfy \(\widehat P_F(g+s)=\widehat P_F(g)\). Its nonnegative
spectrum also satisfies \(\sum_g\sqrt{\widehat P_F(g)}\le2^n\).
Lemma~\ref{low:geometry} and Appendix~\ref{sec:affine-cache} use these
facts.
Every nonzero frequency in \(\mathcal L\) lies outside \(\{0,s\}\).
Because \(G_g\in[0,1]\), Fourier inversion and \eqref{low:isolated} bound
the total nonconstant coefficient mass of \(p_c\) by \(1/8\).
The corresponding coefficients of \(r_c\) are multiplied by \(\beta^c\).
This proves the density bounds and their stated consequences, including
the same calculation at power \(2w\).
\end{proof}

We call \((z_0,\phi_1,\ldots,\phi_k,v_0)\) the chart of a trial, and call the chart transverse when \(\ip{s}{z_0}=1\). The geometry of a trial consists of its restriction \(M\), its chart and its prefix map, which is chosen in the next subsection.

The decoder uses the chart to turn sampled signs into affine coordinates.
For a full row \(Y\) with input coordinates \(X=(\ip{\phi_j}{Y})_{j=1}^k\), called its input,
\[
 \chi_s(Y)=\chi_{v_0}(Y)(-1)^{u\cdot X}.
\]
Thus a row with input \(X=e_j\) gives an observable sign
\(\chi_{v_0}(Y)\) whose correction by \((-1)^{u_j}\) is
\(\chi_s(Y)\). A positive corrected mean lets repeated samples
recover \(u_j\). For \(c\) independent rows with input sum \(x\), the
corrected sign of their XOR has conditional mean \(\gamma_c(x)\) of
Lemma~\ref{low:isolation}, the conditional probability that all \(c\)
rows are clean: a clean row has \(\chi_s=1\), since \(P_F\) is supported
on \(s^\perp\), and the sign of a uniform row is unbiased given its input.

The decoder follows the pattern of the parity algorithm of Blum, Kalai
and Wasserman \cite{BKW}: it combines rows whose inputs sum to a unit
vector and recovers one coordinate from the sign bias of many such
combinations. Each combination is the XOR of one free row from each of
\(2w\) groups and of some pivot rows, which correct the remaining input
difference. Matching full inputs directly
would cost about \(2^{k/2}\) work per output. Keys built from a Hamming
code on blocks of pivot coordinates shrink the key space by the factor
\(2^{m_0}\) per block, at the price of two pivot rows per block;
Lemma~\ref{low:eventual} shows that the net saving
keeps one progress block (Proposition~\ref{low:block}) below \(2^{(1-1/L)n}\) times a polynomial.
Guo, Johansson, and L\"ondahl \cite[Sections~5.4 and~9.3]{GJL}
propose pairing vectors decoded to the same codeword in covering-code
methods for BKW-type LPN algorithms; here the coset-leader differences
select pivot rows whose addition enforces the target input exactly. Selection uses only input
coordinates, after which the character \(\chi_{v_0}\) is evaluated.
Section~\ref{low:decoder} first fixes the decoder's parameters and then
defines it exactly; Appendix~\ref{sec:affine-cache} analyzes it.

\subsection{The affine decoder and its finite parameters}\label{low:decoder}

Here are all remaining public choices. They depend on the fixed noise
promise and on the collision constant \(a_0\) of
Lemma~\ref{coll:explicit-lower}. Write
\begin{equation}\label{low:code-parameters}
\begin{gathered}
 d_{\mathrm{pref}}\coloneqq\lceil\log_2(1024n)\rceil,\quad
 B\coloneqq\left\lfloor\frac{n-d_{\mathrm{pref}}}{L}\right\rfloor,\\
 b_p\coloneqq BL,\quad b_0\coloneqq b_p-2B,\quad c\coloneqq2w+2B,\\
 \vartheta\coloneqq\frac1{16n},\quad h\coloneqq\lceil b_p/4\rceil,\quad
 \xi\coloneqq\left[\frac{L(L-1)}{(L+1)^2}\right]^B,\\
 \rho_0\coloneqq\frac1{16},\quad \ell_0\coloneqq\frac{a_0}{12\cdot1100},\quad
 v\coloneqq\frac{\rho_0^2\ell_0^4}{1024k},\quad B_E\coloneqq\frac{32}{\rho_0^2}.
\end{gathered}
\end{equation}
Require \(n\ge2\), \(w\ge1\), \(B\ge1\), and \(64\le b_p<k\).
These checks precede any formula with a denominator involving \(w\) or \(c\).
Set
\begin{equation}\label{low:reservoirs}
 \begin{split}
 \Lambda&\coloneqq\beta_{\min}^{-2}2^{\lceil k/c\rceil},\\
 M_P&\coloneqq\left\lceil\max\{(n+2)^{20},1024n^2\Lambda/v\}\right\rceil,\\
 q&\coloneqq\text{the least power of two at least }
 \max\{2,(n+2)^{20},2w2^{\lceil k/w\rceil},1024w\Lambda/v\}.
 \end{split}
\end{equation}
The cache of one progress block has \(M_P\) pivot rows and \(2w\) disjoint groups of \(q\)
free rows, all acquired as independent original pair observations.
For each trial the half-uniform replacement and the masks of
\eqref{low:masked-law} are applied afresh, once per original index; the results are the transformed rows of that trial.

Choose an ordered rank-\(b_p\) prefix map on the \(k\) inputs, independently
of the data, by drawing its coordinate functionals uniformly one at a time. Give each functional at most
\(\lceil\log_2(128b_p)\rceil\) attempts to extend the preceding span.
Abort this trial if an extension fails. The total abort probability is at
most \(1/128\), since each attempt succeeds with probability at least
\(1/2\). Conditional on success the prefix is uniform among ordered
rank-\(b_p\) maps: the acceptance probability of each extension depends only on
the dimension of the preceding span.

We now define the decoder on one transformed cache.
Complete the prefix map, by any fixed polynomial-time rule, to input coordinates
\(z\eqqcolon(z_B,z_R)\), so that the prefix \(z_B\) has \(b_p=BL\) bits,
cut into \(B\) blocks of length \(L\).
The binary Hamming code of length \(L\) has as parity-check columns all
nonzero vectors of \(\F_2^{m_0}\). Every vector has a unique expression
\(d+e_I\), where \(d\) is a codeword and the leader \(e_I\) is either
zero or a unit vector; its syndrome determines \(I\). Apply this quantizer separately
in the \(B\) blocks, and denote its codeword output by \(\mathcal Q\).

A descriptor consists of an ordered tuple \(\Pi\) of \(b_p\) distinct pivot
indices, a left free sum \(Y\), which selects one row from each of the
first \(w\) groups, and a right free sum \(Y'\), which selects one row
from each of the other \(w\) groups. The tuple \(\Pi\) supplies input
matrices \(\Pi_B,\Pi_R\). Reject it if \(\Pi_B\) is singular. Otherwise put
\[
 p_z\coloneqq z_B\Pi_B^{-1},\qquad q_z\coloneqq z_R+p_z\Pi_R.
\]
For a prescribed full input target \(x\), the free sums \(Y\) and
\(Y'\) have keys
\begin{equation}\label{aff:keys}
 (\mathcal Q(p_Y),q_Y),\qquad
 (\mathcal Q(p_{Y'}+p_x),q_{Y'}+q_x).
\end{equation}
On a key match let \(d\) be the common codeword output, and write
\(p_Y\eqqcolon d+e_I\), \(p_{Y'}+p_x\eqqcolon d+e_J\).
Retain the match precisely when, in every block, \(I,J\) are distinct
nonzero leaders, and
\begin{equation}\label{aff:guard}
 |A_{\mathrm{sh}}|\ge h,\qquad
 A_{\mathrm{sh}}\coloneqq\{i:(d+e_I)_i=(d+e_J)_i=1\}.
\end{equation}
The descriptor is valid when \(\Pi_B\) is invertible and its keys match
and are retained. The constant \(\xi\) of \eqref{low:code-parameters} is
the fraction of pairs of leader tuples \((I,J)\) that are distinct and
nonzero in every block.
The guard ensures that if two descriptors have fewer than \(h\) common pivot
indices, each has a pivot at a position of its \(A_{\mathrm{sh}}\) that is absent from the other's tuple, which
Lemma~\ref{aff:moments} uses.
A valid descriptor outputs the full row
\begin{equation}\label{aff:output}
 Y_{\mathrm{out}}\coloneqq Y+Y'+
                  \sum_{i\in\operatorname{supp}(e_I+e_J)}\Pi_i.
\end{equation}
It uses exactly \(c\) distinct rows: the \(2w\) free rows and the \(2B\)
pivot rows at the leader positions; the other \(b_0=b_p-2B\) pivot rows
are not in the output. For an input \(z\), \(z_B=p_z\Pi_B\) and \(z_R=q_z+p_z\Pi_R\), and a pivot row \(\Pi_i\) has \(p_{\Pi_i}=e_i\) and \(q_{\Pi_i}=0\). The key match gives \(p_Y+p_{Y'}+p_x=e_I+e_J\), in which the common codeword cancels, and \(q_Y+q_{Y'}=q_x\); so \(p_{Y_{\mathrm{out}}}=p_x\) and \(q_{Y_{\mathrm{out}}}=q_x\). Its prefix is therefore \(x_B\),
and its residual is \(q_x+p_x\Pi_R=x_R\); hence its full input is \(x\).
All selection tests use only inputs. The output sign is read after the
descriptor has been selected.
For a fixed cache, let \(Z_x\) be the fraction of descriptors that are
valid, and let \(T_x\) be the same normalized sum weighted by the
corrected output sign \(\chi_s(Y_{\mathrm{out}})\). Write \(z_x\coloneqq\E Z_x\)
and \(t_x\coloneqq\E T_x\) for their population values at fixed
geometry, and use subscripts \(j\) when \(x=e_j\).
Equation~\eqref{aff:cache-statistics} writes these sums out.

The following rational quantities specify the public tests:
\begin{align}\label{low:guards}
 \tau_h&\coloneqq\min\left\{1,\frac{\binom{b_p}{h}^2}{\binom{M_P}h}\right\},&
 \overline p&\coloneqq\min\{1,2^{n+\lceil n/(2c-1)\rceil-k}\},\nonumber\\
 G_{\mathrm{pref}}&\coloneqq
 \frac{(2^{b_p}-1)(2^n-1)}{(2^k-1)\vartheta},&
 \Delta_{\mathrm{unit}}&\coloneqq(2^{2B+1}-1)(2^{-b_p}+2\overline p).
\end{align}
In addition to \(M_P\ge b_p\), require
\begin{gather}\label{low:finite-tests}
 G_{\mathrm{pref}}\le\frac1{16},\qquad
 \frac{8k\Delta_{\mathrm{unit}}}{\beta_{\min}^c}\le\frac1{16},\qquad
 64\,2^{2(b_p+k)}\beta_{\min}^{-2c}\tau_h\le\frac v2,\\
 E_{\mathrm{low}}\coloneqq r+\left\lceil\frac{k-Bm_0}{2}\right\rceil
                   +2B+2cg_0
       \le n-\lceil n/L\rceil.\nonumber
\end{gather}
Here \(G_{\mathrm{pref}}\) controls the flatness of the prefix law, \(\overline p\) and \(\Delta_{\mathrm{unit}}\) the rank failures, and \(\tau_h\) large pivot overlaps between two descriptors, all in Lemma~\ref{low:geometry}; \(E_{\mathrm{low}}\) is the cost exponent of Proposition~\ref{low:block}. All these tests use known integers and rationals. A failing test invokes
the finite-dimensional fallback of Section~\ref{sec:finite-fallback}.

\begin{lemma}\label{low:geometry}
Under \eqref{low:promise} and the public tests, a complete trial has an
event of probability at least \(2^{-r}/4\), depending only on
its restriction, chart, and prefix, with the following property.
For every fixed geometry in this event the hypotheses of
Lemma~\ref{aff:population} hold simultaneously for the targets
\(e_1,\ldots,e_k\). Their population valid-descriptor masses \(z_j\) and
signed numerators \(t_j\), with target character \(s\), satisfy
\begin{equation}\label{low:population-counts}
 z_j\ge\frac{21}{256}2^{Bm_0-k}\xi,\qquad
 \frac{t_j}{z_j}\ge\frac{\gamma_c(e_j)}4
                         \ge\frac{\beta_{\min}^c}{8}.
\end{equation}
The actual cache quantities \(Z_j,T_j\) have relative variances at most
\(4V_{\mathrm{guard}}\), where
\begin{equation}\label{low:variance}
 V_{\mathrm{guard}}\coloneqq
 81\left[(1+b_p\Lambda/M_P)^{b_p}(1+\Lambda/q)^{2w}-1\right]
 +64\,2^{2(b_p+k)}\beta_{\min}^{-2c}\tau_h\le v.
\end{equation}
\end{lemma}

\begin{proof}
Condition first on \eqref{low:isolated} and the transverse chart of
Lemma~\ref{low:isolation}. A uniform prefix has
\(\sum_{0\ne g\text{ in its frequency space}}\sqrt{\widehat P_F(g)}
\le\vartheta\), except with probability at most \(G_{\mathrm{pref}}\):
each nonzero input frequency is included with probability
\((2^{b_p}-1)/(2^k-1)\), and their total square-root mass is at most \(2^n-1\).
The input-prefix law of one transformed row is therefore pointwise between
\((1-\vartheta)2^{-b_p}\) and \((1+\vartheta)2^{-b_p}\).

Lemma~\ref{aff:population} assumes this flatness, with
\(b_0\vartheta\le1/8\), and, for each target \(e_j\), that its average
rank-failure probability \(\overline\delta(e_j)\) is at most
\(\gamma_c(e_j)/4\). Since \(r_c(0)/r_c(e_j)\le2\) by
\eqref{low:flat-density}, the bound \eqref{aff:rank-average}, proved
after that lemma from Lemma~\ref{aff:subset-entropy}, gives
\(\E\,\overline\delta(e_j)\le\Delta_{\mathrm{unit}}\) over the uniform
prefix.
Markov's inequality and a union over the \(k\) targets show that,
except with probability at most \(1/16\), all these averages are at most
\(\beta_{\min}^c/8\le\gamma_c(e_j)/4\).

Both prefix properties hold with probability at least \(7/8\).
On them Lemma~\ref{aff:population} applies to every target \(e_j\),
since \(b_0\vartheta\le1/16\), and gives
\(z_j\ge\mathcal N2^{-b_p}r_c(e_j)/8\) and
\(t_j/z_j\ge\gamma_c(e_j)/4\), where the number \(\mathcal N\) of
retained branches satisfies
\(\mathcal N2^{-b_p}\ge\tfrac34\,2^{Bm_0}\xi\) by
\eqref{aff:branch-count}. With \(r_c(e_j)\ge7/(8D_X)\) this gives
\eqref{low:population-counts}. Lemma~\ref{aff:moments} bounds both
relative variances, \(\operatorname{Var}(Z_j)/z_j^2\) and
\(\operatorname{Var}(T_j)/t_j^2\), by
\(V_{\mathrm{guard}}\) times the square of
\(\max\{r_c(0)/r_c(e_j),p_c(0)/p_c(e_j)\}\); by \eqref{low:flat-density}
this square is at most \(4\), which gives the variance assertions. The first term in \eqref{low:variance} is at most
\(486v/1024<v/2\), since
\(\Lambda(b_p^2/M_P+2w/q)\le3v/1024\) and
\(e^x-1\le2x\) for \(0\le x\le1\); the overlap test supplies the
other half. Finally the product of restriction, chart, prefix-generation,
and prefix-goodness probabilities is at least
\((7/8)(1/2)(127/128)(7/8)2^{-r}\ge2^{-r}/4\).
\end{proof}

\subsection{Recovering the affine coordinates from one cache}

Fix any good geometry of Lemma~\ref{low:geometry}. Let
\(s_{L,\mathrm{raw}},s_{R,\mathrm{raw}}\) be the collision probabilities
of the two empirical laws of \(w\)-row free input sums.
Since \(q\ge2w2^{\lceil k/w\rceil}\), Lemma~\ref{aff:energy} gives
\(\E(s_{L,\mathrm{raw}}+s_{R,\mathrm{raw}})\le4r_{2w}(0)\), and by Fourier
inversion and Lemma~\ref{low:isolation},
\(r_{2w}(0)=D_X^{-1}\sum_{g\in\mathcal L}\widehat R_F(g)^{2w}\le9/(8D_X)\).
Thus
\[
 \E(s_{L,\mathrm{raw}}+s_{R,\mathrm{raw}})\le\frac9{2D_X}.
\]
Hence the single event
\begin{equation}\label{low:energy-event}
 s_{L,\mathrm{raw}}+s_{R,\mathrm{raw}}\le B_E/D_X
\end{equation}
fails with probability at most \(9/(2B_E)\). For an invertible pivot
tuple \(\Pi\) and a target \(x\), let \(p_L,p_R\) be the laws of the keys
\eqref{aff:keys} of uniform left and right free sums, and call
\[
 e_{\Pi,x}\coloneqq\sum_zp_L(z)^2+\sum_zp_R(z)^2+2\sum_zp_L(z)p_R(z)
\]
the key collision sum. Lemma~\ref{aff:energy} gives
\(e_{\Pi,x}\le2^{Bm_0+1}(s_{L,\mathrm{raw}}+s_{R,\mathrm{raw}})\) on every
cache, so on this event \(e_{\Pi,x}\le2^{Bm_0+1}B_E/D_X\), simultaneously
for every invertible pivot tuple \(\Pi\) and every affine target \(x\).
Use the public certificates
\begin{equation}\label{low:thinning-certificates}
 z_{\min}\coloneqq\frac1{64}2^{Bm_0-k}\xi,\qquad
 K_{\mathrm{use}}\coloneqq128B_E\xi^{-1}.
\end{equation}
On \eqref{low:energy-event}, whenever \(|Z_j-z_j|\le z_j/3\), the bounds \eqref{low:population-counts} imply
\(Z_j\ge z_{\min}\) and
\(e_{\Pi,e_j}\le K_{\mathrm{use}}Z_j\).

Given these certificates for a target \(x\), Lemma~\ref{aff:transport}
supplies, for every positive integer \(L_{\mathrm{late}}\), a finite
sampler that reads only inputs and the public bounds and succeeds with
probability at least \(1-2^{-L_{\mathrm{late}}}\). Its successful
outputs split into early ones, returned by a cache-dependent analysis scale \(H_*\), whose descriptor law
conditional on success by that scale is between \(\ell_0\) and \(\ell_0^{-1}\) times the actual uniform
valid-descriptor law, and late ones, whose probability conditional on success is at most
\(2^{1-L_{\mathrm{late}}}\). The signs of the transformed rows, conditioned on the
full input cache, are independent and have nonnegative posterior means.
Lemma~\ref{aff:sign-transfer} turns these facts into a bound on signs:
for a target \(x\) satisfying the hypotheses of
Lemma~\ref{aff:population}, if \(Z_x\) and \(T_x\) have relative
variances at most \(V_0\), and a
selector that reads only inputs has descriptor weights between \(\ell\)
and \(\ell^{-1}\) times the actual uniform valid law on the intersection
of an input event \(\mathcal E\) with \(|Z_x-z_x|\le z_x/3\), then,
except on caches of probability at most
\(\Prob(\mathcal E^c)+54\ell^{-4}V_0\), this intersection holds and the
selector's corrected sign mean is at least \(\ell(t_x/z_x)/4\).
Apply it to the early outputs for each target \(e_j\), with
\(\ell=\ell_0\), \(V_0=4V_{\mathrm{guard}}\) and \(\mathcal E\) the
event \eqref{low:energy-event}, charged once for all targets; the
paragraph after Lemma~\ref{aff:sign-transfer} checks the certificates on
this intersection. By \eqref{low:population-counts}, all \(k\) early
corrected means are then at least
\(\ell_0\gamma_c(e_j)/16\ge\ell_0\gamma_c(e_j)/32\), and
the required count events and the event \eqref{low:energy-event} hold, except with probability at most
\begin{equation}\label{low:cache-failure}
 216k\ell_0^{-4}V_{\mathrm{guard}}+\frac9{2B_E}
 <\frac{\rho_0^2}{2}.
\end{equation}
These means are taken over the sampler coins, with the complete cache, including its saved signs, held fixed; \eqref{low:cache-failure} bounds the probability of the exceptional caches.

Choose the least positive integer \(L_{\mathrm{late}}\) such that
\(2^{-L_{\mathrm{late}}}\le\ell_0\beta_{\min}^c/512\).
Since the late outputs have conditional probability at most
\(2^{1-L_{\mathrm{late}}}\), they
change a successful corrected sign mean by at most
\(4\cdot2^{-L_{\mathrm{late}}}\).
On the event above, its corrected mean is thus at least
\(\ell_0\beta_{\min}^c/128\).
Represent FAIL by the real number zero. Each such response then has
corrected mean at least
\begin{equation}\label{low:signal}
 \sigma\coloneqq\ell_0\beta_{\min}^c/256.
\end{equation}
For target \(e_j\), the computed output sign is
\(\chi_{v_0}(Y_{\mathrm{out}})\), where \(Y_{\mathrm{out}}\) is the output row \eqref{aff:output}; its corrected sign is obtained by
multiplication by \((-1)^{u_j}\). Take
\[
 N_{\mathrm{read}}\coloneqq
 \left\lceil8\sigma^{-2}\lceil\log_2(8k/\rho_0^2)\rceil\right\rceil
\]
independently seeded responses per target, and set \(\widehat u_j\coloneqq0\)
exactly when their sum is positive, and \(\widehat u_j\coloneqq1\) otherwise. Conditional on the complete cache these responses are
independent. Hoeffding's inequality and a union bound show that every
coordinate is correct except with probability at most \(\rho_0^2/4\).
When every coordinate is correct, the candidate \(v_0+\sum_j\widehat u_j\phi_j\) equals
\(s\).
For each fixed good geometry the complete decoder therefore fails with
probability less than \(\rho_0^2\), averaged over the original cache,
its fresh per-row transformations, and the sampler coins.

\subsection{Repeating restrictions without new measurements}

\begin{lemma}\label{low:reuse}
Let \(0<\rho<1\), let \(\Omega\) be an original random cache, and let a fresh trial choose
geometry \(\Gamma\), independently of \(\Omega\), and further independent
coins. Suppose that a set \(\mathcal G\) of geometries has probability
\(p\ge p_*>0\), and that for each fixed \(\Gamma\in\mathcal G\) a capped
trial fails with probability at most \(\rho^2\), averaging over the
cache and the further coins. Then \(T\) trials with independent coins
on the same cache all fail with probability at most
\(\rho+\exp(-(1-\rho)p_*T)\).
\end{lemma}
\begin{proof}
Let \(Q(\Omega,\Gamma)\) be the failure probability over the further coins.
Tonelli's theorem gives
\(\E_\Omega\E_{\Gamma}[\one_{\mathcal G}Q]\le\rho^2p\).
Except on a set of caches of probability at most \(\rho\), Markov's
inequality bounds the inner expectation by \(\rho p\).
On any remaining cache the success probability of one trial through
\(\mathcal G\) is at least \((1-\rho)p\). The trials are independent
conditional on that original cache, proving the result.
\end{proof}

\begin{proposition}[One low-moment progress block]\label{low:block}
Assume \eqref{low:promise} and the public tests. Fix
\(C\subseteq S^\perp\) with \(S\nsubseteq C\) before acquiring the
block's original data. There is a procedure using polynomially many
original copies and polynomial working storage that emits at most
\(2^{r+6}\) candidate labels. With probability greater than \(7/8\),
at least one candidate belongs to \(S\setminus C\).
Its bit-operation and fair-bit counts on every history are at most
\(2^{E_{\mathrm{low}}}\operatorname{poly}_a(n)\).
If an independent heavy-label test, on one simultaneous event, passes
every ideal stabilizer and passes only labels with \(B_g\ge a/2\),
stopping at the first passing label outside \(C\) gives a label outside \(C\) with \(B_g\ge a/2\)
with probability greater than \(7/8\), conditional on that event.
The resource caps hold also when the moment or subspace hypotheses fail.
\end{proposition}

\begin{proof}
Acquire exactly \(M_P+2wq\) independent original pair rows, at a cost of
\(4(M_P+2wq)\) copies. In each of \(2^{r+6}\) trials choose a new
restriction, chart, and capped random prefix, transform every original row
with fresh replacement and mask coins, and perform the fixed readouts above.
When \(z_0=0\) or prefix generation aborts, this trial returns FAIL.
Use Lemma~\ref{low:reuse} with
\(p_*\coloneqq2^{-r}/4\) and \(\rho\coloneqq\rho_0=1/16\).
The failure probability is at most \(1/16+e^{-15}<1/8\).
An earlier passing proposal is already a heavy label outside \(C\), as asserted.
Fresh acquisition after fixing \(C\) makes the same argument valid
conditional on any history that selected \(C\).

By Lemma~\ref{aff:transport}, one response costs at most
\[
 [K_{\mathrm{use}}+\sqrt{K_{\mathrm{use}}/z_{\min}}]
       \operatorname{poly}_a(n,L_{\mathrm{late}})
 \le \xi^{-1}2^{\lceil(k-Bm_0)/2\rceil}\operatorname{poly}_a(n).
\]
The \(k\) readouts add a factor \(\beta_{\min}^{-2c}\) times a
polynomial. The restriction trials add \(2^r\).
Since \(\xi^{-1}\le2^{2B}\) and
\(\beta_{\min}^{-2c}\le2^{2cg_0}\), these give the stated exponent.
Preparing each transformed cache and evaluating the heavy-label test add
polynomial work per trial. The latter's confidence parameter can be
included in the polynomial when the block is used in the final assembly.
All loops, including failed calls and off-promise trials, have their public
caps. Store only the original cache, one transformed cache, matrices,
counters, and one set of seeds at a time.
\end{proof}

\begin{lemma}\label{low:eventual}
For fixed \(a\) and fixed collision constants, every public test in this
section holds for all sufficiently large \(n\). The cache sizes \(M_P\) and \(q\) are
polynomial in \(n\), and the exponent test gives the block cap
\(2^{(1-1/L)n}\operatorname{poly}_a(n)\).
\end{lemma}
\begin{proof}
The choices in the common setup give
\(w=\Theta_a(n/\log n)\),
\(b_p=n-O_a(\log n)\), \(B=n/L-O_a(\log n)\), and
\(c=2n/L+o(n)\). Since \(w\ge1\) gives \(w\ge n/(2P_0\ell_n)\), we have
\(k/w\le4P_0\ell_n\), so all terms in \eqref{low:reservoirs} are
polynomial whenever the test \(w\ge1\) passes; \(v^{-1}=O_a(n)\).
Since \(b_p\le n-d_{\mathrm{pref}}\), the choice of \(d_{\mathrm{pref}}\) makes \(G_{\mathrm{pref}}\le1/16\).
The second test in \eqref{low:finite-tests} has exponential factor
\(2^{-n+(2+2g_0)n/L+o(n)}\), which tends to zero.
For the overlap test,
\[
 \tau_h\le\binom{b_p}{h}(b_p/M_P)^h
       \le\left(\frac{e b_p^2}{hM_P}\right)^h
       =\exp[-\Omega(n\log n)].
\]
This absorbs its \(\exp[O_a(n)]\) multiplier and the inverse-polynomial
tolerance. Finally, \(r\le a_n+2g_0w+6\) and the even choice
\(m_0=8g_0+8\) give
\[
 E_{\mathrm{low}}
 \le n-2B+6g_0w+\lceil\eta n\rceil+6.
\]
Indeed, \(\lceil(2n-1-Bm_0)/2\rceil=n-Bm_0/2\).
Since \(P_0\ge1024(g_0+1)/\eta\), we have
\(6g_0w\le(3/512)\eta n\). Together with
\(B\ge n/L-d_{\mathrm{pref}}/L-1\), this yields
\[
 E_{\mathrm{low}}
 \le n-\frac{101885}{51200}\frac nL
       +\frac{2d_{\mathrm{pref}}}{L}+9.
\]
For \(n\ge32L+8d_{\mathrm{pref}}\),
\((50685/51200)n/L\ge2d_{\mathrm{pref}}/L+10\), so this is
at most \(n-n/L-1\le n-\lceil n/L\rceil\).
This sufficient condition holds eventually; the algorithm retains the exact
exponent test together with all its other public tests. All integers used in the tests have
polynomial bit length, so the exceptional finite set is handled by testing
the displayed inequalities directly.
\end{proof}

\section{Collisions at a large Fourier moment}\label{sec:high-moment}

We now treat the high-moment promise of
\eqref{eq:moment-split}. When two independent sums of \(w\) working labels are equal, their common
value \(x\) has a law almost supported on \(S^\perp\),
while translation by \(S\) makes \((-1)^{\ip zx}\) take each value
with probability \(1/2\) under that law for every \(z\notin S^\perp\).
The construction below samples such common values from polynomially
many original observations. It returns an isotropic space containing
\(S\) with fixed probability. Independent repetitions provide a list containing a suitable space with high probability; the terminal procedure in
Proposition~\ref{prop:terminal-list} then gives the final containment and purity guarantees.

Throughout this section, \(N=2n\), \(D_V\coloneqq|V|=2^N\), and
\(R,\beta,f,w,A,A_{\min}\) are as in
Section~\ref{sec:moment-framework}. In particular,
\[
 R=\beta P+(1-\beta)U,
 \qquad A=\sum_{g\in V}f(g)^{2w},
 \qquad A_{\min}=2^{\lfloor\eta n\rfloor}.
\]
The algorithm uses the public lower threshold \(A_{\min}\); \(A\)
enters only the analysis.

\subsection{The common-label distribution}

Write \(H_w\coloneqq R^{*w}\) and \(P_w\coloneqq P^{*w}\).
Then
\[
 H_w=\beta^wP_w+(1-\beta^w)U,
 \qquad \zeta\coloneqq\sum_xH_w(x)^2=A/D_V.
\]
Conditional on equality of two independent \(H_w\)-distributed labels,
their common label has law
\begin{equation}\label{high:square-law}
 K_w(x)\coloneqq\frac{H_w(x)^2}{\zeta}.
\end{equation}
Both \(P_w\) and its pointwise square are supported on \(S^\perp\)
and invariant under translation by \(S\). Expanding the square gives
the three nonnegative mixture weights
\begin{equation}\label{high:mixture-weights}
 \frac{\beta^{2w}D_V\|P_w\|_2^2}{A},\qquad
 \frac{2\beta^w(1-\beta^w)}{A},\qquad
 \frac{(1-\beta^w)^2}{A}
\end{equation}
on the normalized square of \(P_w\), on \(P_w\), and on \(U\),
respectively. Their sum is one, since
\(A=\beta^{2w}D_V\|P_w\|_2^2+1-\beta^{2w}\).
Consequently
\begin{equation}\label{high:population-clean}
 K_w(V\setminus S^\perp)\le A^{-1}.
\end{equation}
Moreover, for every \(z\notin S^\perp\), translation by a suitable
\(s\in S\) interchanges the two values of
\(\chi_z(x)=(-1)^{\ip zx}\). Thus
\begin{equation}\label{high:population-spread}
 \Prob_{K_w}[\chi_z(x)=-1]=\tfrac12.
\end{equation}
We next prove the corresponding bounds for the implemented sampler;
its distribution on a fixed cache need not equal \(K_w\).

\subsection{One original cache}

For one output call, acquire \(c_h\coloneqq2w\) disjoint groups of \(q_h\)
independent working rows, using four original copies per row. A left
descriptor chooses one row from each of the first \(w\) groups, and a
right descriptor does the same in the other groups. There are
\(T\coloneqq q_h^w\) descriptors on each side. Their keys are the full
\(N\)-bit XORs, denoted \(Y_u\) and \(Y'_v\). Descriptors are evaluated
on demand.

On the fixed cache define
\begin{align*}
 Z&\coloneqq T^{-2}\sum_{u,v}\one\{Y_u=Y'_v\},\\
 Z_z&\coloneqq T^{-2}\sum_{u,v}\one\{Y_u=Y'_v,\ \chi_z(Y_u)=-1\},\\
 Z_{\rm out}&\coloneqq T^{-2}\sum_{u,v}
          \one\{Y_u=Y'_v,\ Y_u\notin S^\perp\}.
\end{align*}
The last two statistics are used only in the proof. Every individual
descriptor pair uses distinct original rows, so
\begin{equation}\label{high:cache-means}
 \E Z=\zeta,\qquad \E Z_z=\zeta/2\quad(z\notin S^\perp),
 \qquad \E Z_{\rm out}\le D_V^{-1}.
\end{equation}
Indeed, outside \(S^\perp\), \(H_w(x)=(1-\beta^w)/D_V\), which
proves the last bound after averaging over the original cache.

\begin{lemma}[Count and spread on an original cache]\label{high:cache}
Put
\[
 \overline\Lambda\coloneqq2^{\lceil N/c_h\rceil},\qquad
 V_h\coloneqq(1+\overline\Lambda/q_h)^{c_h}-1.
\]
For every fixed \(z\notin S^\perp\),
\begin{equation}\label{high:cache-variance}
 \operatorname{Var}(Z)\le\zeta^2V_h,
 \qquad\operatorname{Var}(Z_z)\le\zeta^2V_h.
\end{equation}
If \(p_L,p_R\) are the empirical full-key laws and
\(S_{\rm CP}\coloneqq\sum_xp_L(x)^2+\sum_xp_R(x)^2\), then
\begin{equation}\label{high:cache-energy}
 \E S_{\rm CP}\le4\zeta
 \quad\text{when}\quad
 q_h\ge2w\,2^{\lceil N/w\rceil}.
\end{equation}
\end{lemma}

\begin{proof}
Two independently uniform descriptor pairs share \(J\) original
indices, where \(J\) has distribution \(\operatorname{Bin}(c_h,1/q_h)\).
For \(J=0\), their indicators in \(Z\) are independent, and so are
their indicators in \(Z_z\). If \(J=j\), the joint equality probability
is exactly
\[
 M(j)\coloneqq D_V^{-2}\sum_{g_1,g_2\in V}
       f(g_1)^{c_h-j}f(g_2)^{c_h-j}f(g_1+g_2)^j.
\]
This follows by Fourier-expanding the two zero-XOR conditions; a
shared row contributes \(f(g_1+g_2)\), and each remaining row contributes
\(f(g_1)\) or \(f(g_2)\). Lemma~\ref{coll:young} implies
\[
 M(j)/\zeta^2\le(D_V/A)^{j/c_h}\le\overline\Lambda^j.
\]
The joint indicator in \(Z_z\) is at most the joint indicator in \(Z\). For overlapping descriptors discard the nonnegative
product of the marginal means. For \(j\ge1\), use
\(\overline\Lambda^j\le(1+\overline\Lambda)^j-1\), and average the
binomial overlap law. This gives both bounds in
\eqref{high:cache-variance}, including repeated descriptors and
overlaps on either side. Equation~\eqref{high:cache-energy} is
Lemma~\ref{coll:group-energy} with the full key space.
\end{proof}

Fix the constants
\begin{equation}\label{high:constants}
 B_h\coloneqq1024,\qquad M_h\coloneqq32+256(B_h+2),\qquad
 \ell_h\coloneqq\frac{a_0}{12M_h},\qquad
 J_h\coloneqq\left\lceil\frac{16M_h}{a_0}\right\rceil,
\end{equation}
where \(a_0\) is the fixed collision constant from
Appendix~\ref{sec:collision}. Choose \(q_h\) as the least power of two
at least
\begin{equation}\label{high:reservoir}
 \max\left\{2,\ 2w\,2^{\lceil N/w\rceil},\
           2048c_h\overline\Lambda\right\}.
\end{equation}
Then \(V_h\le e^{1/2048}-1\le1/1024\). Define the cache event
\[
 \mathcal E_{\rm cache}\coloneqq\{Z\ge\zeta/2,\ S_{\rm CP}\le B_hZ\}.
\]
Chebyshev and Markov give
\begin{equation}\label{high:good-cache}
 \Prob(\mathcal E_{\rm cache}^c)\le4V_h+8/B_h.
\end{equation}

\subsection{The capped common-label sampler}

Assume the public tests \(w\ge1\) and
\(A_{\min}\le D_V/128\) hold. Run the dyadic scales
\begin{equation}\label{high:schedule}
 h=256,512,\ldots,h_{\max},\qquad
 h_{\max}\coloneqq2D_V/A_{\min}.
\end{equation}
The endpoint is a power of two and is at least 256. At each scale run
at most \(J_h\) fresh whole trials of Lemma~\ref{coll:thinning},
with its parameters \(B\), \(H\) and \(\overline F\) equal to \(B_h\), \(h\) and \(2M_hh\).
A whole trial draws fresh four-wise independent arrays of \(h\) left and \(h\) right descriptors (Lemma~\ref{coll:finite-maps}) and runs the capped collision walk of Lemma~\ref{coll:capped-pair} on their compressed keys. The keys are the full labels. Every candidate is checked to be a
cross-side collision of the original keys, after which its common
label is returned. A common label equal to zero is accepted. Return
failure only if the full schedule ends without an accepted pair.
Every trial redraws its tuple maps, key compression, and collision
hash and start.

\begin{lemma}[Implemented spread and contamination]\label{high:one-output}
If \(A\ge A_{\min}\), then for every fixed \(z\notin S^\perp\),
one complete output call satisfies
\begin{align}
 \Prob[\chi_z(\text{output})=-1]&\ge\ell_h/16,
 \label{high:actual-spread}\\
 \Prob[\text{output}\notin S^\perp]&\le6J_h/A_{\min}.
 \label{high:actual-contamination}
\end{align}
Failure is interpreted as output zero. The second estimate holds
for every value of \(A\).
\end{lemma}

\begin{proof}
On \(\mathcal E_{\rm cache}\), let \(h_*\) be the first scale with \(h_*Z\ge1\).
It exists because \(Z\ge A_{\min}/(2D_V)\). If \(h_*=256\), then
\(h_*Z\le256\) since \(Z\le1\); otherwise the previous scale gives
\(h_*Z<2\). Thus the hypotheses of Lemma~\ref{coll:thinning}, with
\(B=B_h\), \(H=h\), \(\mu=Z>0\) and \(s_L+s_R=S_{\rm CP}\le B_hZ\), hold
at every scale \(h\le h_*\), since there \(hZ\le h_*Z\le256\). At that scale success has probability at least
\(a_0/(8M_h)\) per trial. Its \(J_h\) independent attempts fail with
probability at most \(e^{-2}<1/4\).

The schedule truncated after \(h_*\) consequently succeeds with
probability at least \(3/4\). Conditional on success, its descriptor
law is a convex mixture of laws each bounded below by \(\ell_h\)
times the uniform law on matching descriptor pairs: preceding failures give a multiplier
common to all possible outputs of the next trial. The truncated
schedule is used only in the proof; the implemented schedule is fixed
by \eqref{high:schedule}.

For a fixed \(z\notin S^\perp\), intersect \(\mathcal E_{\rm cache}\) with
\(Z\le2\zeta\) and \(Z_z\ge\zeta/4\). By
\eqref{high:cache-variance}, its probability is at least
\[
 1-21V_h-8/B_h\ge1-29/1024>15/16.
\]
On this event the common label of a uniform matching pair has odd probability
\(Z_z/Z\ge1/8\). The truncated schedule therefore has odd-output
probability at least \((3/4)\ell_h/8\). Later outputs of the complete
schedule can only add to this probability. Averaging over the
original cache proves
\[
 \Prob[\chi_z(\text{output})=-1]
 \ge(15/16)(3/4)\ell_h/8\ge\ell_h/16.
\]

For contamination we use the upper bound on each output atom in
\eqref{coll:descriptor-atoms}, which holds on every cache and at every
scale, regardless of their collision probabilities. For a fixed cache a
whole trial at scale \(h\) returns outside \(S^\perp\) with probability
at most \((3h/2)Z_{\rm out}\). Stopping at the first success is bounded
by the union of all potential trial outputs. Since
\(\sum_hh<2h_{\max}\), the conditional bound is
\(3J_hh_{\max}Z_{\rm out}\). Take its expectation and use
\eqref{high:cache-means} to obtain
\[
 \Prob[\text{output}\notin S^\perp]
 \le3J_hh_{\max}/D_V=6J_h/A_{\min}.
\]
This bound averages over every original cache.
\end{proof}

\subsection{Independent outputs and the radical}

Set
\begin{equation}\label{high:output-count}
 R_h\coloneqq\left\lceil\frac{16}{\ell_h}(N+5)\right\rceil.
\end{equation}
Run \(R_h\) complete calls using mutually independent original caches
and fresh classical randomness. Insert each returned label into a
span \(C_h\), with failure treated as zero, and discard its original
cache after the call. Return \(\rad(C_h)=C_h\cap C_h^\perp\).

\begin{proposition}[The high-moment radical]\label{high:radical}
Suppose the public tests
\begin{equation}\label{high:guards}
 n\ge2P_0\lceil\log_2(n+2)\rceil,\qquad
 A_{\min}\le D_V/128,\qquad
 \frac{6R_hJ_h}{A_{\min}}\le\frac1{32}
\end{equation}
hold. The procedure just defined returns an isotropic subspace on
every execution. Under \(A\ge A_{\min}\), it contains \(S\) with
probability at least \(15/16\). Its original-copy count and workspace
are polynomial in \(n\), and its bit work and fair bits are bounded
on every execution by
\begin{equation}\label{high:resources}
 2^{(1-\eta/2)n}\operatorname{poly}_a(n).
\end{equation}
The tests in \eqref{high:guards} eventually hold at fixed \(a\), and
their exceptional dimensions are independent of the requested final
confidence.
\end{proposition}

\begin{proof}
For fixed \(z\notin S^\perp\), Lemma~\ref{high:one-output} and
independence of the complete calls imply
\[
 \Prob[\chi_z\text{ equals }+1\text{ on every output}]
 \le e^{-\ell_hR_h/16}\le e^{-(N+5)}.
\]
Union over at most \(2^N\) characters. The resulting failure
probability is at most \(2^Ne^{-(N+5)}<1/32\). Outside that failure
event, \(C_h^\perp\subseteq S^\perp\), equivalently \(S\subseteq C_h\).
Separately, \eqref{high:actual-contamination} and
\eqref{high:guards} give
\(\Prob[C_h\not\subseteq S^\perp]\le1/32\). Hence with probability
at least \(15/16\),
\[
 S\subseteq C_h\subseteq S^\perp.
\]
On this event \(S\subseteq C_h\cap C_h^\perp\). The radical of any
subspace of \(V\) is isotropic, so the output has that
property even on all remaining histories and when the high-moment
promise is false.

There are exactly \(4R_hc_hq_h\) original state copies in this
construction. The first test in \eqref{high:guards} implies
\(w\ge n/(2P_0\lceil\log_2(n+2)\rceil)\), and therefore
\(N/w\le4P_0\lceil\log_2(n+2)\rceil\). Thus \(q_h\), \(c_hq_h\)
and \(R_h\) are polynomial in \(n\), with constants and degrees
depending only on \(a\) and the fixed collision constant.

One scale costs \(J_h\sqrt h\) times a polynomial in
\(n,w,q_h\), by the deterministic cap in
Lemma~\ref{coll:thinning}. The geometric sum over
\eqref{high:schedule} is \(O(\sqrt{D_V/A_{\min}})\). Hence all calls
cost
\[
 R_hJ_h\sqrt{D_V/A_{\min}}\operatorname{poly}_a(n)
 \le\sqrt2\,2^{(1-\eta/2)n}\operatorname{poly}_a(n).
\]
This counts cache preparation, tuple evaluation, full-key checks,
seed generation, and output reconstruction, including failed
trials. Stored-row access can be implemented by scanning the
polynomial-size cache, which changes only the polynomial factor.
The cache stores \(O(Nc_hq_h)\) label bits; the current sampler seed
and counters have polynomial length, and a span basis uses
\(O(N^2)\) bits. Caches are processed sequentially. Fair bits are
generated sequentially and each trial seed is discarded after use.

Finally, \(n/\log(n+2)\to\infty\), \(\eta<1\), and
\(A_{\min}=2^{\lfloor\eta n\rfloor}\) grows exponentially, while
\(R_h=O_a(n)\) and \(J_h\) is constant. These observations verify
eventual satisfaction of all three public tests. All confidence
parameters in this proposition are the fixed values \(1/32\).
\end{proof}

\section{The complete learner}\label{sec:assembly}
We now combine the two moment constructions. Their intermediate spaces serve different purposes: a heavy-label span records constraints on the ideal stabilizers, its radical supplies a commuting candidate, as for the batch spans in \cite[Section~II.B.1]{TS}, and the final space is grown only through the observed-state purity verifier. Keeping those steps separate permits the same terminal procedure to use every returned radical.

The complete learner receives \(n,t,a\) and the precision integers
\(e,b\) of Section~\ref{mem:model}; for \(t=n\) it replaces \(t\) by \(n-1\) internally before all
subsequent tests and calls. If \(n-t\ge\kappa n\), it runs the learner of
Proposition~\ref{prop:linear-gap} with gap \(\kappa\), which either
searches all \(4^n\) labels with the verifier or applies
Proposition~\ref{prop:terminal-list} to \(b+6\) radicals of
Lemma~\ref{mem:lem-radical}. Otherwise it runs the public tests of
Sections~\ref{sec:low-moment} and~\ref{sec:high-moment}, using the
fallback of Section~\ref{sec:finite-fallback} if one fails; if all pass,
it builds one low-moment radical and \(b+6\) high-moment radicals
(Sections~\ref{sec:low-span} and~\ref{sec:moment-list}) and applies
Proposition~\ref{prop:terminal-list} to the list of all of them. Every
branch ends with the verifier of Section~\ref{mem:verification}.

\subsection{From low-moment proposals to a containing radical}\label{sec:low-span}
Recall that \(N=2n\), and use the dyadic confidence \(\delta=2^{-b}\) of Section~\ref{mem:model}. Independently of every proposal cache, acquire a holdout of
\[
 M_{\mathrm{heavy}}\coloneqq\left\lceil32a^{-2}(N+b+8)\right\rceil
\]
original raw Bell labels. For every nonzero \(g\), their empirical character mean estimates a signed quantity of magnitude \(\alpha B_g\). Hoeffding's inequality and a union over the \(2^N\) characters give an event \(\mathcal E_{\mathrm{heavy}}\), of failure probability at most \(\delta/128\), on which all these means have error at most \(a/4\). Character means are computed as queried, from the same polynomial-size cache.

Initialize a linear space \(C\coloneqq0\). A proposed label passes the heavy-label test precisely when it lies outside \(C\) and its empirical absolute character mean is at least \(3a/4\). On \(\mathcal E_{\mathrm{heavy}}\), every ideal stabilizer outside \(C\) passes, since \(\alpha\ge a\). Every passing label satisfies \(\alpha B_g\ge a/2\), hence \(B_g\ge a/2>0\), using \(\alpha\le1\), and \(g\in S^\perp\). Thus \(C\subseteq S^\perp\) remains invariant.

At each fixed-\(C\) stage, run at most
\[
 J_{\mathrm{low}}\coloneqq b+\lceil\log_2(N+1)\rceil+6
\]
independent complete blocks of Proposition~\ref{low:block}. Acquire each block's cache after the current \(C\) has been fixed. When a label passes, enlarge \(C\) by that label and restart the stage. Stop after a stage in which no label passes, or after \(N\) rank increases. The latter cap applies on every history.

Under the low-moment promise, at a stage with \(S\nsubseteq C\), each block yields a passing label with probability greater than \(7/8\), conditional on the past and \(\mathcal E_{\mathrm{heavy}}\). This is the last assertion of Proposition~\ref{low:block}, applied to the heavy-label test: couple each block with its complete proposal sequence, which continues after the first passing label; on \(\mathcal E_{\mathrm{heavy}}\) a recovered missing stabilizer passes the test, and every earlier passing label already enlarges \(C\). Fresh block caches then bound the failure probability of such a stage by \(2^{-J_{\mathrm{low}}}\). A union over at most \(N+1\) stages costs at most \(\delta/64\). On \(\mathcal E_{\mathrm{heavy}}\), \(N\) rank increases are impossible because \(\dim S^\perp=N-\dim S\le N-1\). Consequently, except with probability at most \(\delta/128+\delta/64\),
\[
 S\subseteq C\subseteq S^\perp.
\]
Return \(R_{\mathrm{low}}\coloneqq\rad(C)=C\cap C^\perp\). This space is isotropic on every history, and the displayed inclusions imply \(S\subseteq R_{\mathrm{low}}\). The heavy-label test has established a containing candidate radical; final purity is controlled by the separate procedure below.

\subsection{A list that covers both moment regimes}\label{sec:moment-list}
Run
\[
 K_{\mathrm{pre}}\coloneqq b+6
\]
independent high-moment procedures from Proposition~\ref{high:radical}, each with the fixed internal confidence values \(1/32\) stated there. Every output call within such a procedure uses its own original cache, as required there. Keep all returned radicals, including zero spaces. Under the high-moment promise, the output of each procedure contains \(S\) with probability at least \(15/16\), so the probability that none does is at most \(16^{-K_{\mathrm{pre}}}\le\delta/64\).

Form a list of these \(K_{\mathrm{pre}}\) radicals together with \(R_{\mathrm{low}}\). Every member is isotropic on every history. Since the two promises in \eqref{eq:moment-split} cover every state, the list contains a member including \(S\), except with probability at most
\[
 \delta/128+\delta/64+\delta/64.
\]
This bound adds the two regime-specific failure bounds, so it holds in either regime. Proposition~\ref{prop:terminal-list} uses only the existence of a good member; its index can remain unknown.

\subsection{Public finite-dimensional fallback}\label{sec:finite-fallback}
The low- and high-moment procedures impose public tests, which involve only public parameters and fixed base confidence. Their proofs show that all tests hold eventually at each fixed noise promise. The tests include the low exponent bound and the high contamination bound. Their failures therefore occur in a finite set of dimensions depending only on fixed algorithm constants. Final confidence is amplified by fresh complete procedures, so this exceptional set is independent of the requested \(\delta\).

If any test fails, acquire the verifier cache of Section~\ref{mem:verification} and scan all \(4^n\) labels for eligible extensions, approving them by the verifier rule. Restart after an approval; stop after an empty scan or at rank \(n\). Cap the number of scans by \(n+1\). On \(\mathcal E_{\mathrm{ver}}\), a missing ideal stabilizer is eligible and has zero loss. Hence an empty scan implies containment. The isotropy and total-loss conclusions follow from the same verifier invariant. Copies and storage are polynomial. The factor \(4^n\) in time is absorbed into a fixed constant on this finite set of dimensions.

\subsection{Proof of the main theorem and resource bounds}
For \(n-t\ge\kappa n\), Proposition~\ref{prop:linear-gap} applies with \(c=\kappa\) and reduced noise promise \(a_c=\min\{a,\kappa\}/2\). Its polynomial depends only on the input \(a\), because \(\kappa\) does. In the complementary range, perform the public tests, using the fallback if needed. Otherwise construct the candidate list and apply Proposition~\ref{prop:terminal-list} with an independent verifier cache of size \eqref{mem:verifier-count}. The empirical approval threshold and the uniform estimation error are both \(\varepsilon_0/(2n)\). Their sum is the allowed true loss \(\varepsilon_0/n\) per approval.

The failure bounds for the list, the verifier cache and the terminal list procedure sum to \(5\delta/128+\delta/128+\delta/64=\delta/16<\delta\). Thus the resulting \(H\) contains \(S\) and has total observed-state loss at most \(\varepsilon_0\le\varepsilon\), with the required probability.

For resources, Proposition~\ref{low:block} and the exponent test in \eqref{low:finite-tests} give a deterministic per-block time and fair-bit cap
\[
 2^{(1-1/L)n}\operatorname{poly}_a(n,b),
\]
where \(b\) enters through the queries to the heavy-label test.
There are at most \((2n+1)J_{\mathrm{low}}\) complete blocks, each using a fresh polynomial cache. Within a block all restriction trials reuse that cache, storing one transformed cache at a time. The holdout contributes only polynomial copies and storage.

Each base high procedure has cap
\[
 \sqrt{4^n/A_{\min}}\operatorname{poly}_a(n)
 \le\sqrt2\,2^{(1-\eta/2)n}\operatorname{poly}_a(n),
\]
and polynomial original copies and workspace. There are \(K_{\mathrm{pre}}=O(b)\) such procedures. Their sample caches can be discarded sequentially; the stored list of radical bases has polynomial size. In the small-gap range \(t>(1-\kappa)n\), both exponents \((1-1/L)n\) and \((1-\eta/2)n\) lie below \(t\), because \(\kappa=\eta/4\) is smaller than \(1/L\) and \(\eta/2\). The terminal list procedure has cap \(2^t\operatorname{poly}(n,b,1/\varepsilon)\) directly.

All sampling, transformations, exact key tests, seed generation, capped retries, failed executions, matrix operations, heavy-label tests and counters are included in these bounds. Seeds are generated from sequential fair bits and evaluated in polynomial workspace. Amplifying confidence adds polynomial factors in \(b\), while all base constants and public tests remain fixed. The fallback costs at most \((n+1)4^n\) polynomial-time tests; on its finite set of dimensions this is at most \(2^t\) times an \(a\)-dependent constant and a polynomial. This proves Theorem~\ref{thm:all-gap} for \(t<n\).

For \(t=n\), run this same algorithm with internal input \(t^{\prime}\coloneqq n-1\). If \(S\ne\{0\}\), its nullity is at most \(n-1\), so the preceding proof applies. If \(S=\{0\}\), containment is automatic. Every branch, including the fixed-gap branch, returns the output of either the terminal list procedure or the finite-dimensional fallback. Eligibility preserves isotropy on every history. On the simultaneous verifier event, at most \(n\) accepted extensions each lose at most \(\varepsilon_0/n\), giving the required loss bound independently of finder success. This event fails with probability at most \(\delta/128\). All public resource caps apply on off-promise histories as well, with \(2^{t^{\prime}}\le2^t\). The reduced noise promise in the fixed-gap branch remains valid because \(\alpha\ge a\ge a_c\). This proves the endpoint and completes the theorem.

\section{Conclusion}\label{sec:conclusion}
For every fixed promised bound on global depolarizing noise below one, polynomially many copies and polynomial working storage suffice for unsigned stabilizer containment with bounded additive purity loss. Classical processing takes \(2^t\) times a polynomial, using Bell measurements alone. The guarantee covers every \(0\le t\le n\), including a single promised independent stabilizer, and answers the threshold question of Tang and Steckmann \cite[Section~X]{TS} for global depolarizing noise.

The proof uses a fixed-gap feature filter adapted from Chen, De, and Vijayaraghavan \cite{CDV}, together with two overlapping moment regimes for small gaps. In the low-moment regime a polynomial cache supports rare isolation events through affine decoding; in the high-moment regime, collisions found with a primitive based on the analysis of Lyu and Zhu \cite{LZ} yield labels that are sufficiently spread and rarely anticommute with an ideal stabilizer. The radicals of the resulting label spans feed a common observed-state verifier. As in the finder--verifier organization of Tang and Steckmann \cite{TS}, separating candidate construction from verification allows polynomial measurement resources even when the classical proposal search has exponential length.

Polynomial degrees, sufficient cutoffs and finite-range constants depend on the noise promise and can be very large: even for the promise \(\lambda\le1/2\), which gives the smallest cutoffs, evaluating the first test in \eqref{high:guards} and the cutoff \(N_f\) of Section~\ref{sec:linear-gap} at the promise parameter \(a=(1-1/2)^2=1/4\) and the fixed gap \(c=\kappa\) shows that the high-moment construction is used only for \(n\ge2.59\times10^{22}\) and the fixed-gap filter only for \(n\ge2^{321}\); below these sizes, the corresponding branches are replaced by exhaustive searches over all \(4^n\) Pauli labels, whose exponential work factor is bounded by a constant depending on the promise. Useful end-to-end numerical bounds and an implementation remain quantitative directions for further work. The dependence on a fixed promise also leaves room for bounds polynomial in \(1/(1-\lambda)\) as well, which would allow \(\lambda\) to approach one as \(n\) grows; by Proposition~\ref{prop:bell-noise-lower-bound}, for learners using only standard Bell measurements with failure probability at most \(1/4\), such copy bounds must grow at least as \((1-\lambda)^{-4}\).

Complete depolarization gives a sharp identifiability boundary for this objective. At \(\lambda=1\), the states \(|0\rangle\) and \(|+\rangle\) both yield \(I/2\), while their ideal stabilizers \(Z\) and \(X\) anticommute. Their containment events are therefore disjoint for an isotropic output. Every learner has failure probability at least \(1/2\) on one of these two inputs, even with unrestricted measurements.

General Pauli channels, coherent errors and measurement noise require model-specific extensions. The optimal dependence of the running time on the nullity is also open: the weak-noise polynomial-time guarantee of Tang and Steckmann \cite[Theorems~II.2 and~VIII.5]{TS} and the present \(2^t\) bound raise their question \cite[Section~X]{TS} of whether one algorithm can interpolate between these running times as the noise strength varies.

\appendix
\section{Collision sampling with a pseudorandom hash family}\label{sec:collision}

This appendix constructs the capped collision sampler used by both moment
branches. The input is an implicitly represented array: its entries can be
recomputed from a stored cache and a short seed. We first obtain a work cap that
keeps a lower bound on the output probability of each collision, and then transfer that
bound through fresh random restrictions of the implicit array.

\subsection{A work cap with a lower bound for every collision}

Let \(M\ge1024\) be a power of four, let
\(\mathcal A:[M]\to[m]\) be fixed with \(m\le M^{O(1)}\), and let
\[
 F\coloneqq\sum_{i,j\in[M]}\one\{\mathcal A(i)=\mathcal A(j)\}
\]
be its second frequency moment \cite[Section~3]{LZ}, the number of ordered index pairs with equal values.
A collision is an unordered pair \(e=\{u,v\}\) of distinct indices
with \(\mathcal A(u)=\mathcal A(v)\). Array evaluation is deterministic and its bit work
and workspace are charged below.

We use the hash family of Chen, Jin, Williams, and Wu \cite{CJWW}
in the version analyzed by Lyu and Zhu \cite[Section~4.1]{LZ}, with depth \(d_0\coloneqq\frac12\log_2M\)
and independence parameter \(20\log_2M\). It consists of independent
levels, each taking value zero with probability \(1/2\), and each
nonzero value in \([M]\) with probability \(1/(2M)\). The first nonzero
level defines \(h\); if all levels vanish, \(h\) returns a termination
symbol. This family has a fair-bit seed of length
\(O(\log^3 M)\), polynomial-time evaluation, and workspace polynomial in \(\log M\). Choose an independent uniform start \(x_0\in[M]\), and let
\(O\) be the distinct vertices reached by iterating \(h\circ\mathcal A\).
Write \(R_O\coloneqq|O|\) and \(E_e\coloneqq\{u,v\in O\}\).

The proof uses the recursive walk tree of \cite[Definition~4]{LZ}.
Its indices are vectors of nonnegative integers in \(\mathbb N^{d_0}\),
with root zero. The parent of a nonzero index is obtained by decreasing
its first nonzero coordinate by one. The width of an index is its largest
coordinate. For an index set \(J\), let \(\mathcal P(J)\) be the union
of its root-to-index paths, including their endpoints, and assign it
weight \(2^{1-|\mathcal P(J)|}\).
The source's extended walk is an auxiliary experiment: it bounds each
coordinate by \(\tau\coloneqq5\log_2M\) and replaces a repeated
level-hash query on these paths by an independent transition that terminates the level with probability
\(1/2\) and otherwise chooses a uniform vertex \cite[Algorithm~2]{LZ}.
For \(j\) prescribed, distinct, \(\tau\)-bounded indices and prescribed
vertices, its joint probability is
\(M^{-j}2^{1-|\mathcal P(J)|}\), provided each level hash is
\(j\tau\)-wise independent \cite[Lemma~5]{LZ}.
The index sum obeys \cite[Lemma~9]{LZ}
\[
 \sum_{|J|=j}2^{1-|\mathcal P(J)|}\le2^{jd_0}.
\]
These identities allow the two following proofs to compare the actual
orbit with the auxiliary walk.

Set \(a_0\coloneqq1/61440\). The next bound makes the constant in the collision guarantee of \cite[Lemma~4]{LZ} explicit for the collisions used here.
\begin{lemma}[Explicit pair probability]\label{coll:explicit-lower}
For every collision \(e\) of \(\mathcal A\) in the preceding experiment,
\begin{equation}\label{coll:published-lower}
 \Prob(E_e)\ge a_0/F.
\end{equation}
\end{lemma}
\begin{proof}
Write \(d\coloneqq d_0\), \(\tau=10d\) and \(p_e\coloneqq\Prob(E_e)\). We use the finite estimates in the proof of \cite[Lemma~4]{LZ}, whose Section~4.4 bounds \(p_e\) below by \(SE_1-SE_2-SE_3\); we retain the main term \(SE_1\) before its asymptotic simplification, and the three subtracted terms below are its bounds on \(SE_2\) and \(SE_3\), with \(24\cdot2^{4(\gamma+1)}=384\cdot2^{4\gamma}\) and \(6\cdot2^{3(\gamma+1)}=48\cdot2^{3\gamma}\). For every integer \(2\le\gamma\le d\), these give
\begin{align}
 p_e\ge{}&\frac{(2-2^{-\tau})^{2\gamma}-2^\gamma}{M^2}
 -\frac{384F2^{4\gamma}}{M^4}
 -\frac{48\sqrt F\,2^{3\gamma}}{M^3}\notag\\
 &-\frac{6\gamma(\tau+1)^2 2^{3\gamma-\tau}}{M^2}.
 \label{coll:finite-lower}
\end{align}
To obtain the main term, restrict the first \(\gamma\) index coordinates
to be at most \(\tau\) and set all higher coordinates to zero; the
sampled hash retains all \(d\) levels. The sum of the one-index path
weights is \((2-2^{-\tau})^\gamma\). The sum over ordered pairs of
indices, including the diagonal, is at least its square, and deleting
the diagonal costs at most \(2^\gamma\). For a collision of \(\mathcal A\),
the standard-walk index events are disjoint
\cite[Section~4.2.2, equation~(4)]{LZ}.

Lemmas~6--8 of that source subtract events where a collision or a width
cutoff invalidates this comparison. Its Lemma~9 and Corollary~1 bound
the ordered index sums for the collision terms: four endpoints contribute
\(4!2^4=384\), and three contribute \(3!2^3=48\). The value class
of a specified endpoint has size at most \(\sqrt F\).
For the width term there are \(\gamma\) coordinate choices and at most
\((\tau+1)^2\) choices of the deleted endpoint coordinates. Deleting
the width coordinate saves a factor \(2^{-\tau}\), and the
three-index sum gives \(6\cdot2^{3\gamma}\). These give the three
subtracted terms in \eqref{coll:finite-lower}.

The endpoint identities expose at most \(4\tau=20\log_2M\) distinct
probes at each level, so the selected exact \(4\tau\)-wise independence
of each level suffices. The dimension reduction for the width sum is
valid for \(\gamma-1\ge1\). The retained finite leading term needs only \(\gamma\ge2\); Lyu and Zhu invoke \(\gamma\ge5\) only for the final simplification in their equation~(7).

Put \(y\coloneqq\sqrt F/M\). The two disjoint events that the start is one endpoint and its successor is the other show
\[
 p_e\ge\frac{2(1-2^{-d})}{M^2}\ge\frac{31}{16M^2},
\]
since \(d\ge5\). If \(y\ge1/330\), then
\(Fp_e\ge31/1742400>1/61440\).

Otherwise \(F\ge M\) implies \(M>330^2\), so the power-of-four condition gives \(d\ge9\). Choose the least integer \(\gamma\) such that \(x\coloneqq2^\gamma y\ge1/165\). Then
\[
 2\le\gamma\le d,\qquad 1/165\le x<2/165,
\]
because \(2^dy=\sqrt{F/M}\ge1\). Multiplying \eqref{coll:finite-lower} by \(F\) and using Bernoulli's inequality yields
\[
 Fp_e\ge\phi(x)-E_*,\qquad
 \phi(x)\coloneqq\tfrac34x^2-48x^3-384x^4,
\]
where \(E_*\coloneqq\gamma2^{-\tau}x^2+6\gamma(\tau+1)^2 2^{\gamma-\tau}x^2\) satisfies
\[
 E_*\le d2^{-10d}+6d(10d+1)^2 2^{-9d}
 \le7d(10d+1)^2 2^{-9d}<2^{-62}.
\]
The final inequality holds at \(d=9\), since \(7\cdot9\cdot91^2=521703<2^{19}\), and its upper bound decreases for subsequent integers \(d\). On \([1/165,2/165]\), \(\phi''(x)=3/2-288x-4608x^2<0\), so its minimum is at an endpoint. The two endpoint values are
\[
 \phi(1/165)=\frac{16153}{988267500},\qquad
 \phi(2/165)=\frac{16228}{988267500}.
\]
Thus \(Fp_e\ge16153/988267500-2^{-62}>1/61440\); the last comparison follows from
\[
 \frac{16153}{988267500}-\frac1{61440}
 =\frac{69547}{1011985920000}>2^{-24}.
\]
\end{proof}

\begin{lemma}[Endpoint bounds]\label{coll:endpoints}
For the experiment above,
\[
 \Prob(E_e)\le\frac3M,
 \qquad
 \E\bigl[(R_O+1)\one_{E_e}\bigr]\le\frac{20}{\sqrt M}.
\]
\end{lemma}

\begin{proof}
We spell out the consequences of the extended-walk bounds that are
needed in addition to \eqref{coll:published-lower}.
The standard and extended walks agree through any prescribed indices if no
collision of \(\mathcal A\) or width cutoff occurs before their last index
\cite[Section~4.3.2]{LZ}. The width-failure event \(\mathcal B\)
satisfies
\[
 \Prob(\mathcal B)\le\tau2^{d_0-\tau}\le M^{-3};
\]
this follows by summing the first width-cutoff contribution over its
coordinate, as in the proof of \cite[Lemma~3]{LZ}.

The orbit of one start contains at most one collision of \(\mathcal A\). Indeed,
when a value repeats, its successor has already occurred, so its later
endpoint is the last new vertex. Consequently, on \(E_e\), there is no
earlier collision of \(\mathcal A\) before the later endpoint of \(e\). On
\(\mathcal B^c\), the coupling therefore applies through both endpoints,
including the later one. There are two assignments to a two-index set,
so
\[
 \Prob(E_e)\le
 \frac2{M^2}\,2^{2d_0}+M^{-3}\le\frac3M.
\]
If \(z\notin\{u,v\}\) is also visited, it is visited before the later
endpoint of \(e\). The same argument, with at most six assignments and
\(3\tau\le20\log_2M\), gives
\[
 \Prob(E_e,\ z\in O,\ \mathcal B^c)
 \le\frac6{M^3}\,2^{3d_0}=\frac6{M^{3/2}}.
\]
Sum over \(z\), charging \(\mathcal B\) once by \(R_O\le M\).
Then
\[
 \E[R_O\one_{E_e}]
 \le2\Prob(E_e)+\frac{6(M-2)}{M^{3/2}}+M^{-2}.
\]
Adding \(\Prob(E_e)\le3/M\) proves the asserted bound for
\(M\ge1024\). The coupling has been used only to bound a standard-walk
event from above by an extended-walk event.
\end{proof}

The pair can be recovered using a constant number of orbit positions; the walk and this recovery follow the collision-finding procedure of Beame, Clifford, and Machmouchi \cite{BCM}.
Run Floyd's cycle-finding
algorithm on \(h\circ\mathcal A\), checking the termination symbol at every
evaluation. Termination precludes a collision of \(\mathcal A\). Otherwise let
\(m_{\rm tail}\) and \(m_{\rm cyc}\) be the tail and cycle lengths. If
\(m_{\rm tail}=0\), the orbit is a cycle, in which no vertex has two
predecessors, so it contains no collision; return failure. If
\(m_{\rm tail}>0\), reconstruct the two predecessors
\[
 u'\coloneqq(h\circ\mathcal A)^{m_{\rm tail}-1}(x_0),\qquad
 v'\coloneqq(h\circ\mathcal A)^{m_{\rm tail}+m_{\rm cyc}-1}(x_0)
\]
of the cycle entry, and return them precisely when \(\mathcal A(u')=\mathcal A(v')\).
This recovers \(e\) exactly on \(E_e\): a functional orbit with a tail
has only this one vertex with two distinct predecessors within its
reachable set. Cycle detection, length recovery and reconstruction use
\(O(R_O+1)\) evaluations and \(O(\log M)\) additional bits.
Choose a computable bound \(Q\), polynomial in \(\log M\) and the
parameters bounding the cost of evaluating one array entry, such that its total bit work \(W\),
including seed generation, validation and any specified output
reconstruction, obeys \(W\le Q(R_O+1)\) on every history.

\begin{lemma}[Capped pair selection]\label{coll:capped-pair}
Given a public \(\overline F>0\), stop the preceding trial at the
deterministic work limit
\begin{equation}\label{coll:pair-cap}
 W_{\max}\coloneqq\left\lceil\frac{40Q\overline F}{a_0\sqrt M}\right\rceil
  =\left\lceil\frac{2457600Q\overline F}{\sqrt M}\right\rceil,
\end{equation}
and release a pair only after reconstruction and validation have
finished. For every collision \(e\), its output subprobability is at
most \(3/M\), for every array. If \(F\le\overline F\), it also satisfies
\begin{equation}\label{coll:pair-atoms}
 \frac1{122880F}=\frac{a_0}{2F}\le p_e^{\rm cap}\le\frac3M.
\end{equation}
All work, fair bits and workspace are bounded on every history.
\end{lemma}

\begin{proof}
Lemma~\ref{coll:endpoints} gives
\(\E[W\one_{E_e}]\le20Q/\sqrt M\). Thus
\[
 \Prob(E_e,W>W_{\max})\le\frac{a_0}{2\overline F}
 \le\frac{a_0}{2F}.
\]
Subtract this from \eqref{coll:published-lower}. Truncation only removes
outputs, proving the unconditional upper bound. The work limit is
explicit, and the stored seed, cycle state and counters have polynomial
length. If \(W_{\max}\) is too short even for initialization, the trial returns
failure within the same cap.
\end{proof}

In particular, accepting a computable subset of collisions cannot
increase any output atom. Independent repetitions of the same capped
trial, stopped at the first acceptable output, preserve its conditional
output law: every successful atom is multiplied by the same geometric
sum of preceding failure probabilities. The conditional output-law identity holds also when running time
depends on the output.

\subsection{Fresh restrictions of an implicit array}

\begin{lemma}[Finite tuple maps and key compression]\label{coll:finite-maps}
Let \(H\ge256\) and \(q\ge2\) be powers of two, and let \(w\ge1\).
There is an exact four-wise independent uniform map
\([H]\to[q]^w\) with a seed and evaluation cost polynomial in
\(\log H\) and \(w\log q\). Two independent maps give left and right
arrays of \(H\) descriptors each. For keys in \(\F_2^k\), these arrays
can be compressed and padded to length \(M\), where
\[
 1024\le M,\qquad 2H\le M<8H,
\]
with a polynomial-size alphabet. Every false key equality has
probability \(M^{-3}\); all true equalities are preserved. Every seed
uses ordinary independent fair bits.
\end{lemma}

\begin{proof}
Identify \([H]\) with \(\F_2^{\log_2H}\).
For each of the \(w\log_2q\) output bits choose independent fair
coefficients of a multilinear polynomial of degree at most three in
the \(\log_2H\) input bits. This takes
\(w\log_2q\sum_{j=0}^3\binom{\log_2H}{j}\) bits. At any at most four distinct
points, evaluation has full row rank: for each selected point, multiply
at most three affine linear functions that equal one there and vanish
at the respective other points, and multilinearize. The resulting
polynomial gives the corresponding unit evaluation vector. Uniform
coefficients therefore give independent uniform evaluations.

Let \(M\) be the least power of four at least \(2H\).
Choose a fresh uniformly random binary matrix defining
\(K:\F_2^k\to\F_2^{3\log_2M}\). Genuine entries receive the tagged
key \((0,Kx)\), and each padding position receives its own distinct
tag-one key. The alphabet has at most \(M^3+M\) values. For distinct
\(x,y\), the vector \(K(x-y)\) is uniform, proving the false-equality
probability. Tuple expansion, matrix evaluation and padding are
recomputed from polynomially many stored bits. In every application,
an accepted pair is checked against its original keys.
\end{proof}

Consider two deterministic lists of \(T\coloneqq q^w\) descriptors, accessed
implicitly. Let \(p_L,p_R\) be the key laws of uniform descriptors, and
write
\[
 \mu\coloneqq\sum_xp_L(x)p_R(x),\qquad
 s_L\coloneqq\sum_xp_L(x)^2,\qquad s_R\coloneqq\sum_xp_R(x)^2.
\]
We call an original descriptor pair matching when its left and right keys
agree. A whole trial draws fresh tuple maps and compression as above,
then the hash and start of Lemma~\ref{coll:capped-pair}; it accepts only
cross-side pairs with equal original keys. Repeated original
descriptors and zero keys are retained.

\begin{lemma}[Transport to original descriptors]\label{coll:thinning}
Suppose \(\mu>0\), \(s_L+s_R\le B\mu\), \(B\ge2\), and
\(H\mu\le256\). Set
\[
 M_B\coloneqq32+256(B+2),\qquad \overline F\coloneqq2M_BH,
 \qquad \ell_B\coloneqq\frac{a_0}{12M_B}.
\]
Use the cap \eqref{coll:pair-cap} with this \(\overline F\).
For every fixed matching original descriptor pair \(e\), the whole-trial
output subprobability, averaged over all its fresh seeds, satisfies
\begin{equation}\label{coll:descriptor-atoms}
 \frac{a_0H}{8M_BT^2}\le a_e(H)\le\frac{3H}{2T^2}.
\end{equation}
The upper bound holds for all key laws \(p_L,p_R\).
Under the stated assumptions, success has probability at least
\(a_0H\mu/(8M_B)\), and its conditional descriptor law lies between
\(\ell_B\) and \(\ell_B^{-1}\) times the uniform law on matching original
descriptor pairs. Each trial costs
\(\sqrt H\) times a polynomial in \(B\) and the entry and seed parameters,
with the fixed \(a_0\) absorbed into its constant,
on every history.
\end{lemma}

\begin{proof}
Fix a matching original pair \(e=(u,v)\) with common key \(x\), and
prescribe its occurrence in a particular left and right slot. This has
probability \(T^{-2}\). Before compression, the conditional expected
ordered collision count is exactly
\begin{align*}
 \E[F_{\rm orig}\mid u,v\text{ in the prescribed slots}]
 ={}&M+2+(H-1)(H-2)(s_L+s_R)\\
 &+2(H-1)^2\mu
       +4(H-1)(p_L(x)+p_R(x)).
\end{align*}
The terms count diagonals, the forced cross-pair, remaining pairs on
the same side, remaining cross-pairs, and pairs with one forced
endpoint. Four-wise independence suffices also after prescribing one
slot on each side. Compression adds at most
\(4H^2/M^3\le1/M\) in expectation, independently of the prescriptions.
Using \(M<8H\), \(p_L(x)+p_R(x)\le2\), and \(H\mu\le256\), the result
is at most \(M_BH\). Hence
\[
 \Prob(F\le2M_BH\mid\text{the prescribed occurrence})\ge\tfrac12.
\]
For each such occurrence, Lemma~\ref{coll:capped-pair} gives conditional
output probability at least
\((1/2)a_0/(2\overline F)=a_0/(8M_BH)\).
For every array it gives the upper probability
\(3/M\le3/(2H)\). Sum over the \(H^2\) cross-slot pairs and multiply
by \(T^{-2}\). The indexed-pair output events are disjoint, even when
the original descriptor repeats, so this proves
\eqref{coll:descriptor-atoms}.

There are \(T^2\mu\) matching original pairs. Summing their atoms bounds
the success probability between \(a_0H\mu/(8M_B)\) and
\(3H\mu/2\). Dividing an individual atom by these two bounds gives
the asserted density comparison. Finally
\(\overline F=2M_BH\) and \(M\asymp H\) in
\eqref{coll:pair-cap} give the work bound, independently of whether
the hypotheses on the key laws hold.
\end{proof}

Fresh tuple maps are part of each whole trial, as required by the
seed averaging in \eqref{coll:descriptor-atoms}.
A deterministic function of the returned
pair, such as its common full key or the XOR of the corresponding
rows, can be appended with its evaluation cost included in \(Q\).

\subsection{Moments of grouped caches}

The following bounds control the actual cache by accounting for every
descriptor overlap. All norms in this subsection use unnormalized
counting measure.

\begin{lemma}[A finite-group moment inequality]\label{coll:young}
Let \(G\) be a finite elementary abelian group of order \(d\),
\(b:G\to[0,\infty)\), \(c\ge1\), and \(0\le j\le c\). Put
\(B_u\coloneqq\sum_x b(x)^u\), with \(B_0=d\). Then
\begin{equation}\label{coll:young-bound}
 \sum_{x,y}b(x)^{c-j}b(y)^{c-j}b(x+y)^j
 \le d^{j/c}B_c^{2-j/c}.
\end{equation}
\end{lemma}

\begin{proof}
For \(j\le c/2\), Young's convolution inequality with input exponent
\(c/(c-j)\) and output exponent \(c/(c-2j)\), followed by H\"older,
bounds the left side by
\[
 B_c^{2(c-j)/c}B_{c/2}^{2j/c}
 \le d^{j/c}B_c^{2-j/c}.
\]
The usual infinity-norm conventions include both endpoints. For
\(j\ge c/2\), the convolution infinity bound and the \(\ell_1\) norm
of \(b^j\) give \(B_{2(c-j)}B_j\). Apply
\(B_u\le d^{1-u/c}B_c^{u/c}\) to both factors. The zero function is
handled directly, with empty products interpreted as one.
\end{proof}

\begin{lemma}[Collision probability of grouped sums]\label{coll:group-energy}
Let a distribution on \(\F_2^k\) have nonnegative Fourier transform
\(f\), and let each of \(w\) independent groups contain \(q\)
independent rows. Write \(p_L\) for the law, on the fixed
cache, of the XOR obtained by uniformly choosing one row per group.
For \(d\coloneqq2^k\), \(A\coloneqq\sum_g f(g)^{2w}\), and \(\zeta\coloneqq A/d\),
\begin{align}
 \E\sum_xp_L(x)^2
 &=d^{-1}\sum_g\bigl[(1-q^{-1})f(g)^2+q^{-1}\bigr]^w,
 \label{coll:energy-identity}\\
 \frac{\E\sum_xp_L(x)^2}{\zeta}
 &\le\bigl[1+(d^{1/w}-1)/q\bigr]^w.
 \label{coll:energy-bound}
\end{align}
In particular, two disjoint left and right caches satisfy
\(\E(s_L+s_R)\le4\zeta\) when
\(q\ge2w\,2^{\lceil k/w\rceil}\).
\end{lemma}

\begin{proof}
An empirical row character has second moment
\((1-q^{-1})f(g)^2+q^{-1}\). Multiply over the independent groups
and apply Parseval to obtain \eqref{coll:energy-identity}.
In its binomial expansion the term with \(j\) shared-row factors
contains \(\sum_g f(g)^{2w-2j}\). H\"older bounds its ratio to \(A\)
by \((d/A)^{j/w}\le d^{j/w}\), since \(f(0)=1\).
Summing proves \eqref{coll:energy-bound}. The stated choice of \(q\)
makes this at most \(e^{1/2}<2\) on each side.
\end{proof}

\section{Affine equations from Hamming-coded keys}\label{sec:affine-cache}

This appendix analyzes the affine decoder defined in Section~\ref{low:decoder}.
Fix a masked clean law \(P_F\), a target \(s\in F\cap S\), and a rank-\(k\)
input projection \(X\) whose frequency space excludes \(s\).
Here a row is a transformed row of one trial in Section~\ref{sec:low-moment}, with population law \(R_F=\beta P_F+(1-\beta)U\),
\(\beta\ge\beta_{\min}\). Fourier transforms on \(\F_2^k\) use the characters \(y\mapsto(-1)^{g\cdot y}\). In this appendix write
\(f(g)\coloneqq\widehat{(R_F)_X}(g)\),
\(G(g)\coloneqq\widehat{(P_F)_X}(g)\), \(D_X=2^k\),
\(r_j\coloneqq(R_F)_X^{*j}\), and \(p_j\coloneqq(P_F)_X^{*j}\).
The lift of \(g\) is the frequency of \(V\) whose character takes the value \((-1)^{g\cdot Y_X}\) at every row \(Y\); in Section~\ref{sec:low-moment} it is \(\sum_jg_j\phi_j\), so \(f(g)=\widehat R_F(\sum_jg_j\phi_j)\) and \(G(g)=\widehat P_F(\sum_jg_j\phi_j)\).
Thus \(f\ge0\), \(\sum_g\sqrt{f(g)}\le2^n\), and the full-space
Fourier coefficient at \(s\) plus the lift of \(g\) equals \(\beta G(g)\).
Conditional on the input of a row \(Y\), its sign \(\chi_s(Y)\) has mean
\begin{equation}\label{aff:posterior}
 \gamma(Y_X)\coloneqq\frac{\beta(P_F)_X(Y_X)}
 {\beta(P_F)_X(Y_X)+(1-\beta)/D_X}\in[0,1].
\end{equation}
Indeed the clean row is in \(s^\perp\), whereas under the full uniform
law this character is unbiased given the input. Signs of distinct rows
remain independent conditional on all row inputs.

\subsection{Branches and the affine equations}

Use the parameters \(b_p=BL\), \(L=2^{m_0}-1\ge31\),
\(b_0=b_p-2B\), \(c=2w+2B\), and
\(h=\lceil b_p/4\rceil\), with \(b_p\ge64\) and \(w\ge1\), and the
prefix coordinates \(z=(z_B,z_R)\), quantizer \(\mathcal Q\), keys
\eqref{aff:keys}, retention rule \eqref{aff:guard} and output row
\eqref{aff:output} of Section~\ref{low:decoder}.

A branch is a fixed triple \((d,I,J)\) with distinct nonzero leaders in every block, satisfying \eqref{aff:guard}.
Partition the pivot positions
into shared, left-only, right-only, and unused positions according to
\(((d+e_I)_i,(d+e_J)_i)\). Let \(U_{\mathrm{sh}}\) be the sum of shared pivot inputs,
\(U_{\mathrm{left}}\) the left free sum plus left-only pivots, and \(U_{\mathrm{right}}\) the right
free sum plus right-only pivots. The exact branch equations are
\begin{equation}\label{aff:branch-equations}
 U_{\mathrm{left}}+U_{\mathrm{right}}=x,\qquad (U_{\mathrm{sh}}+U_{\mathrm{left}})_B=0.
\end{equation}
The effective sign in \eqref{aff:output} uses the \(c\) rows of
\(U_{\mathrm{left}}+U_{\mathrm{right}}\); shared pivot signs cancel.

Let \(\mathcal N\) be the number of retained branches, and let
\(K_{\mathrm{code}}\coloneqq2^{B(L-m_0)}\),
\(J_{\mathrm{lead}}\coloneqq[L(L-1)]^B\). Then
\begin{equation}\label{aff:branch-count}
 \frac34K_{\mathrm{code}}J_{\mathrm{lead}}
 \le\mathcal N\le K_{\mathrm{code}}J_{\mathrm{lead}},\qquad
 2^{-b_p}K_{\mathrm{code}}J_{\mathrm{lead}}=2^{Bm_0}\xi.
\end{equation}
To see the lower bound, a uniform codeword has pairwise independent,
unbiased coordinates. Its dual simplex code has no nonzero word of weight
one or two, proving this independence also across the independent blocks.
Its weight has mean \(b_p/2\) and variance \(b_p/4\), so the probability of
weight below \(3b_p/8\) is at most \(16/b_p\). Deleting the \(2B\le b_p/8\)
leader positions leaves at least \(b_p/4\) shared positions. This holds for
every fixed leader pair and proves \eqref{aff:branch-count}.

\subsection{Conditional subset sums and the population valid mass}

\begin{lemma}\label{aff:subset-entropy}
Let \(c\ge2\) independent input rows have nonnegative Fourier transform
\(f\), with \(f(0)=1\) and \(\sum_g\sqrt{f(g)}\le2^n\).
Condition their total sum to be \(x\), where \(r_c(x)>0\).
Every nonempty proper subset sum has maximum atom at most
\[
 \frac{r_c(0)}{r_c(x)}
 \min\{1,2^{2cn/(2c-1)-k}\}.
\]
\end{lemma}
\begin{proof}
For a subset of size \(j\), its conditional mass at \(z\) is
\(r_j(z)r_{c-j}(x+z)/r_c(x)\).
Nonnegative Fourier coefficients imply \(r_i(z)\le r_i(0)\).
Put \(v_g\coloneqq\sqrt{f(g)}\) and \(S_u\coloneqq\sum_gv_g^u\).
H\"older interpolation gives, for \(1\le u\le2c\),
\[
 S_u\le S_1^{(2c-u)/(2c-1)}S_{2c}^{(u-1)/(2c-1)}.
\]
Thus
\[
 \frac{r_j(0)r_{c-j}(0)}{r_c(0)}
 =\frac{S_{2j}S_{2c-2j}}{D_XS_{2c}}
 \le\frac{S_1^{2c/(2c-1)}}{D_XS_{2c}^{1/(2c-1)}}
 \le2^{2cn/(2c-1)-k}.
\]
The same ratio is a conditional probability for the zero target and is
therefore at most one.
\end{proof}

\begin{lemma}\label{aff:population}
Suppose that the prefix law of one row is pointwise between
\((1-\vartheta)2^{-b_p}\) and \((1+\vartheta)2^{-b_p}\), with
\(b_0\vartheta\le1/8\). For each retained branch let \(\delta_{\mathrm{br}}(x)\)
be the conditional probability, given \(U_{\mathrm{left}}+U_{\mathrm{right}}=x\), that its \(2B\)
effective pivot prefixes together with \((U_{\mathrm{left}})_B\) are dependent.
Let \(\overline\delta(x)\) average these probabilities over the fixed
retained branches. If
\(\overline\delta(x)\le\gamma_c(x)/4\), where
\(\gamma_c(x)=\beta^cp_c(x)/r_c(x)>0\), then the population valid
mass \(z_x\) and signed numerator \(t_x\) satisfy
\begin{equation}\label{aff:population-bounds}
 z_x\ge\mathcal N2^{-b_p}r_c(x)/8,\qquad
 t_x\ge\mathcal N2^{-b_p}\beta^cp_c(x)/8,\qquad
 t_x/z_x\ge\gamma_c(x)/4.
\end{equation}
\end{lemma}
\begin{proof}
Fix the effective rows. A shared pivot remains by \eqref{aff:guard}.
If the indicated \(2B+1\) prefixes are dependent, full pivot rank and
the second equation of \eqref{aff:branch-equations} cannot both hold.
If they are independent, the remaining \(b_0\) pivot prefixes complete a
basis while their specified nonempty shared sum equals \((U_{\mathrm{left}})_B\)
with probability between
\[
 \pi_-\coloneqq2^{-b_p}\rho_{b_0-1}(1-\vartheta)^{b_0},
 \qquad \pi_+\coloneqq2^{-b_p}\rho_{b_0-1}(1+\vartheta)^{b_0},
 \qquad \rho_{b_0-1}\coloneqq\prod_{i=1}^{b_0-1}(1-2^{-i}).
\]
For uniform columns this follows by an invertible change on the remaining
column indices that makes their specified sum the first column. Fixing
that column costs \(2^{-b_p}\); the others complete the independent
\(2B+1\) columns with probability \(\rho_{b_0-1}\).
For the actual columns, multiply the probability of each complete tuple
by its product density bounds \((1\pm\vartheta)^{b_0}\).

The effective-row posterior sign product lies in \([0,1]\), by
\eqref{aff:posterior}. Its expectation times the full-sum indicator is
\(\beta^cp_c(x)\): only the all-clean term contributes to the signed
convolution. Removing rank failures therefore costs at most
\(r_c(x)\delta_{\mathrm{br}}(x)\) in the signed numerator.
Summing over branches yields
\[
 \begin{split}
 z_x&\ge\mathcal N\pi_-r_c(x)(1-\overline\delta(x)),\qquad
 z_x\le\mathcal N\pi_+r_c(x),\\
 t_x&\ge\mathcal N\pi_-[\beta^cp_c(x)-r_c(x)\overline\delta(x)].
 \end{split}
\]
Now \(\rho_{b_0-1}\ge1/4\),
\((1-\vartheta)^{b_0}\ge7/8\), and \(\pi_-/\pi_+\ge1/2\).
Together with \(\overline\delta\le\gamma_c/4\le1/4\), these prove
the three weaker bounds in \eqref{aff:population-bounds}.
\end{proof}

We bound the rank-failure probability on average over a random prefix. Suppose that
the prefix map is uniform among ordered rank-\(b_p\) maps and independent
of the rows, and that \(r_c(0)/r_c(x)\le2\). Fix a retained branch and
condition on \(U_{\mathrm{left}}+U_{\mathrm{right}}=x\).
Since \(2cn/(2c-1)\le n+\lceil n/(2c-1)\rceil\),
Lemma~\ref{aff:subset-entropy} bounds by \(2\overline p\), with
\(\overline p\) from \eqref{low:guards}, every atom of the XOR of each
nonempty proper subset of the \(c\) effective rows.
Every nonzero relation among the prefix of \(U_{\mathrm{left}}\) and the
prefixes of the \(2B\) effective pivots, namely the left-only and right-only
pivots, is the prefix of such a subset XOR: the subset
either uses only effective pivots or includes the left free group and
omits the right free group; both free groups are nonempty. The relation
therefore holds only if this XOR is zero, with conditional probability
at most \(2\overline p\), or if the prefix map kills it; a uniform
rank-\(b_p\) prefix kills a fixed nonzero vector with probability at
most \(2^{-b_p}\). Summing over the \(2^{2B+1}-1\) relations and
averaging over the retained branches gives, with
\(\Delta_{\mathrm{unit}}\) from \eqref{low:guards},
\begin{equation}\label{aff:rank-average}
 \E_{\mathrm{prefix}}\,\overline\delta(x)\le\Delta_{\mathrm{unit}}.
\end{equation}

\subsection{Moments of the actual cache}

There are \(T_P\coloneqq M_P(M_P-1)\cdots(M_P-b_p+1)\) ordered distinct pivot tuples and
\(T_{\mathrm{leaf}}\coloneqq q^w\) leaves on either side; a leaf chooses one free row from each of \(w\) groups, like a free sum of Section~\ref{low:decoder} or a one-sided descriptor of Appendix~\ref{sec:collision}. A descriptor \(\omega\) selects one
tuple and one leaf on each side. For a branch \(d_{\mathrm{br}}\), let
\(I_{\omega,d_{\mathrm{br}}}\) indicate its complete valid event, including
rank and \eqref{aff:branch-equations}, and let
\(E_{\omega,d_{\mathrm{br}}}\) be its effective original-index set. Define
\begin{equation}\label{aff:cache-statistics}
 \begin{split}
 Z_x&\coloneqq\frac1{T_PT_{\mathrm{leaf}}^2}\sum_{\omega,d_{\mathrm{br}}}I_{\omega,d_{\mathrm{br}}},\\
 T_x&\coloneqq\frac1{T_PT_{\mathrm{leaf}}^2}\sum_{\omega,d_{\mathrm{br}}}
 I_{\omega,d_{\mathrm{br}}}
 \prod_{i\in E_{\omega,d_{\mathrm{br}}}}\chi_s(Y_i).
 \end{split}
\end{equation}
At most one branch contributes for each descriptor, by uniqueness of
Hamming decoding. These averages are over unpadded tuples; the padded
draw used by the sampler will be accounted for separately.

\begin{lemma}\label{aff:moments}
Under the hypotheses of Lemma~\ref{aff:population}, put
\[
 A_x\coloneqq\max\{r_c(0)/r_c(x),p_c(0)/p_c(x)\},\qquad
 \Lambda=\beta_{\min}^{-2}2^{\lceil k/c\rceil}.
\]
For the independent pivot and free pools,
\[
 \frac{\operatorname{Var}(Z_x)}{z_x^2},\quad
 \frac{\operatorname{Var}(T_x)}{t_x^2}
 \ \le A_x^2\left\{
 81[(1+b_p\Lambda/M_P)^{b_p}(1+\Lambda/q)^{2w}-1]
 +64\,2^{2(b_p+k)}\beta_{\min}^{-2c}\tau_h\right\}.
\]
\end{lemma}
\begin{proof}
Expand the variances over pairs of fixed branches and descriptors.
For two independent uniform descriptors, their number \(J\) of common
pivot indices has the hypergeometric law with parameters \((M_P,b_p,b_p)\).
Their number \(F_0\) of common free indices is independent of \(J\) and
has law \(\operatorname{Bin}(2w,1/q)\). If \(J=F_0=0\), the complete
original-index sets are disjoint and covariance is exactly zero.
The signed branch expectations and signed joint expectations are
nonnegative by conditioning on all inputs and using \eqref{aff:posterior}.
For overlapping sets, covariance is consequently bounded by the joint
expectation.

When \(J<h\), each branch has a shared pivot absent from the other's
entire tuple. Choose one such pivot in each branch. These two indices
are distinct, absent from both effective supports, and independent of all
other inputs. Drop the rank predicates. Conditional on the other rows,
integrating these two private pivots bounds the two prefix equations by
\((1+\vartheta)^2 2^{-2b_p}\). For the signed statistic first average
the signs given the inputs: the remaining product of posterior means is
nonnegative and independent of the two private pivot inputs, so the same
bound applies.

If the effective supports overlap in \(j\le J+F_0\) rows, the remaining
two full-input moments are
\begin{align}\label{aff:overlap-moments}
 M_Z^x(j)&\coloneqq D_X^{-2}\sum_{g_1,g_2}
 (-1)^{(g_1+g_2)\cdot x}
 f(g_1)^{c-j}f(g_2)^{c-j}f(g_1+g_2)^j,\\
 M_T^x(j)&\coloneqq D_X^{-2}\sum_{g_1,g_2}
 (-1)^{(g_1+g_2)\cdot x}
 [\beta G(g_1)]^{c-j}[\beta G(g_2)]^{c-j}f(g_1+g_2)^j.
 \nonumber
\end{align}
Their probability definitions are nonnegative. Dropping the character
bounds each by its zero-target value. Lemma~\ref{coll:young} gives
\[
 \frac{M_Z^0(j)}{r_c(0)^2}\le D_X^{j/c},\qquad
 \frac{M_T^0(j)}{[\beta^cp_c(0)]^2}
                     \le\beta^{-2j}D_X^{j/c}.
\]
For the second inequality use \(f\le G=\beta^{-1}(\beta G)\) in the
shared-row factor and \(\sum_g(\beta G(g))^c\ge\beta^c\).
Comparing with \eqref{aff:population-bounds}, summing over branch pairs,
and using \(64(1+\vartheta)^2\le81\), bounds the small-overlap
contribution to either relative variance by
\(81A_x^2\E[\one_{J+F_0>0}\Lambda^{J+F_0}]\).

The hypergeometric factorial moments satisfy
\[
 \E\binom Jj=\frac{\binom{b_p}{j}^2}{\binom{M_P}j}
 \le\binom{b_p}{j}(b_p/M_P)^j.
\]
For an integer \(m>0\), \(\Lambda^m\le(1+\Lambda)^m-1\).
The two overlap laws therefore bound this expectation by
\((1+b_p\Lambda/M_P)^{b_p}(1+\Lambda/q)^{2w}-1\).
For \(J\ge h\), bound each branch-pair joint expectation by one.
The same factorial-moment identity gives \(\Prob[J\ge h]\le\tau_h\).
Use \(r_c(x)\ge2^{-k}/A_x\) and
\(\beta^cp_c(x)\ge\beta_{\min}^c2^{-k}/A_x\) in
\eqref{aff:population-bounds}. This yields the stated remaining term.
\end{proof}

\subsection{Key collision sums and finite sampling}

For a fixed input cache and invertible pivot tuple, let \(p_L,p_R\) be
the exact key laws of uniform leaves under \eqref{aff:keys}. As in Section~\ref{sec:low-moment}, put
\[
 e_{\Pi,x}\coloneqq\sum_zp_L(z)^2+\sum_zp_R(z)^2+2\sum_zp_L(z)p_R(z).
\]
This includes collisions that fail the leader or shared-pivot tests.

\begin{lemma}\label{aff:energy}
For every input cache, invertible pivot tuple, and target,
\[
 e_{\Pi,x}\le2^{Bm_0+1}
               (s_{L,\mathrm{raw}}+s_{R,\mathrm{raw}}).
\]
If \(q\ge2w2^{\lceil k/w\rceil}\), the two raw leaf-input collision
probabilities satisfy
\(\E(s_{L,\mathrm{raw}}+s_{R,\mathrm{raw}})\le4r_{2w}(0)\).
\end{lemma}
\begin{proof}
The coordinate change \(z\mapsto(p_z,q_z)\) is bijective. Each key
fiber has exactly \((L+1)^B=2^{Bm_0}\) inputs, since each codeword has
that many choices of block leaders. The right target translation preserves
fiber sizes. Cauchy--Schwarz within the fibers bounds each key self-collision
probability by \(2^{Bm_0}\) times its raw input self-collision probability.
The cross term is at most the sum of the two key self-collision probabilities.
This proves the deterministic assertion. The expectation follows from
Lemma~\ref{coll:group-energy} applied to the input law; its spectrum is
nonnegative, \(f(0)=1\), and its zero-sum mass is \(r_{2w}(0)\).
\end{proof}

\begin{lemma}\label{aff:transport}
Fix an input cache and a target \(x\). Suppose public rationals
\(K_{\mathrm{use}}\ge1\), \(z_{\min}>0\) certify
\begin{equation}\label{aff:certificates}
 Z_x\ge z_{\min},\qquad e_{\Pi,x}\le K_{\mathrm{use}}Z_x
 \quad\hbox{for every invertible tuple }\Pi.
\end{equation}
For any positive integer \(L_{\mathrm{late}}\), there is a finite
input-only sampler with success probability at least
\(1-2^{-L_{\mathrm{late}}}\). There is a scale \(H_*\), determined by the cache, such that,
conditional on success by scale \(H_*\), the output descriptor has density between \(\ell_0\) and
\(\ell_0^{-1}\) relative to the uniform valid law in \eqref{aff:cache-statistics}.
Conditional on success, the probability of an output after scale \(H_*\) is at most
\(2^{1-L_{\mathrm{late}}}\), so these later outputs change a bounded sign mean by at
most \(2^{2-L_{\mathrm{late}}}\).
On every cache and execution its bit-operation and fair-bit counts are
\[
 [K_{\mathrm{use}}+\sqrt{K_{\mathrm{use}}/z_{\min}}]
\operatorname{poly}(n,q,M_P,L_{\mathrm{late}},
                    \log(2K_{\mathrm{use}}/z_{\min})),
\]
with polynomial working storage, including the saved rows. The fixed
collision constants are absorbed in this polynomial.
\end{lemma}
\begin{proof}
Draw a binary integer from a range of size
\(R_P\coloneqq2^{\lceil\log_2T_P\rceil}\). If it is at least \(T_P\), fail;
otherwise unrank it as an ordered distinct pivot tuple. Repeated quotient
and remainder operations in radices \(M_P,M_P-1,\ldots,M_P-b_p+1\)
give a bijection and polynomial bit work. Reject singular tuples.
Each ordered distinct tuple has probability exactly \(1/R_P\), with
\(T_P\le R_P<2T_P\).

At a dyadic scale \(H\ge256\), form two implicit arrays of \(H\)
leaves using fresh four-wise uniform tuple maps, then compress and pad
their keys as in Lemma~\ref{coll:finite-maps}. Let \(M_{\mathrm{arr}}\)
be the resulting array length, so \(2H\le M_{\mathrm{arr}}<8H\).
Run the capped trial of Lemma~\ref{coll:capped-pair} with
\(\overline F\coloneqq2\cdot1100H\). Before releasing the row
\eqref{aff:output}, check that the pair is cross-side and passes the
original full-key, leader and shared-pivot tests.
Every trial uses fresh tuple, compression, and collision seeds.

Fix a valid original descriptor and prescribed left and right array slots.
After fixing its pivot tuple, its placement probability is \(T_{\mathrm{leaf}}^{-2}\).
If its common key is \(z\), four-wise independence gives, as in the proof of Lemma~\ref{coll:thinning}, the conditional
expectation of the uncompressed ordered collision count
\[
 M_{\mathrm{arr}}+2+(H-1)(H-2)\Bigl(\sum p_L^2+\sum p_R^2\Bigr)
 +2(H-1)^2\sum p_Lp_R+4(H-1)(p_L(z)+p_R(z)).
\]
Compression contributes at most one in expectation. Thus when
\(He_{\Pi,x}\le1024\), the conditional expectation of the full collision
count is at most
\(M_{\mathrm{arr}}+2+H^2e_{\Pi,x}+8H+1\le1100H\).
Conditional Markov's inequality gives probability at least \(1/2\) that
the actual collision count is at most \(\overline F\).
The two bounds on each pair output probability in Lemma~\ref{coll:capped-pair}, summed over the
\(H^2\) cross-slot pairs, therefore give for this original descriptor
the whole-trial atom bounds
\begin{equation}\label{aff:descriptor-atoms}
 \frac{a_0H}{8\cdot1100R_PT_{\mathrm{leaf}}^2}
 \le a_\omega(H)\le\frac{3H}{2R_PT_{\mathrm{leaf}}^2}.
\end{equation}
The upper bound holds on every cache and scale. Indexed-pair output
events are disjoint even when one original descriptor occurs repeatedly.
Their normalized density bounds are \(\ell_0=a_0/(12\cdot1100)\)
and \(\ell_0^{-1}\); summing the lower atoms gives success probability
at least \(a_0HZ_x/(16\cdot1100)\).

Implement scales \(H=256,512,\ldots\) through the first scale at least
\(\max\{256,1/(K_{\mathrm{use}}z_{\min})\}\).
At each scale run at most
\[
 J\coloneqq\left\lceil(16\cdot1100K_{\mathrm{use}}/a_0)
                                      L_{\mathrm{late}}\right\rceil
\]
independent whole trials, stopping at the first valid output.
For the analysis let \(H_*\) be the first scale with
\(H_*K_{\mathrm{use}}Z_x\ge1\).
At every scale through \(H_*\), \(He_{\Pi,x}\le1024\): if
\(H_*=256\), use \(e_{\Pi,x}\le4\); otherwise use
\(H_*K_{\mathrm{use}}Z_x<2\).
By \(H_*\) the failure probability is at most
\(e^{-L_{\mathrm{late}}}\le2^{-L_{\mathrm{late}}}\).
The successful law of the schedule truncated there is a convex mixture
of the laws in \eqref{aff:descriptor-atoms}; all prior-failure factors
are common to the descriptors of a given trial. The implemented schedule's later
output probability is at most \(2^{-L_{\mathrm{late}}}\). Dividing by
success probability at least \(1-2^{-L_{\mathrm{late}}}\) proves the
claimed conditional late mass and sign bound.

One whole trial costs \(\sqrt H\) times a polynomial in the logarithms
of the displayed parameters, by the work cap of Lemma~\ref{coll:capped-pair}. Summing
\(\sqrt H\) geometrically and multiplying by \(J\) gives
\(O([K_{\mathrm{use}}+\sqrt{K_{\mathrm{use}}/z_{\min}}]
L_{\mathrm{late}})\) times the entry-work polynomial.
All scales and counters have bit length polynomial in the
parameters of Section~\ref{sec:low-moment}. Rows can be retrieved by
scanning the polynomial original cache; seeds are generated
sequentially and only the current seeds are stored.
\end{proof}

\subsection{Transferring the signal to the selected row signs}

\begin{lemma}\label{aff:sign-transfer}
Fix a target satisfying Lemma~\ref{aff:population}, and suppose
\(\operatorname{Var}(Z_x)/z_x^2\le V_0\) and
\(\operatorname{Var}(T_x)/t_x^2\le V_0\).
On the intersection of an input-cache event \(\mathcal E\) and
\(|Z_x-z_x|\le z_x/3\), suppose an input-only successful selector has descriptor weights between
\(\ell\) and \(\ell^{-1}\) times the actual uniform valid law,
where \(0<\ell\le1\).
Except with probability at most
\(\Prob(\mathcal E^c)+54\ell^{-4}V_0\), the count satisfies
\(|Z_x-z_x|\le z_x/3\) and the selector's corrected sign mean on the
complete fixed cache is at least \(\ell(t_x/z_x)/4\).
For a polynomial list of targets a common input event is charged once,
and the remaining bounds are summed over the targets.
\end{lemma}
\begin{proof}
Let \(\mathcal X\) denote all row inputs and put
\(T_{\mathcal X}\coloneqq\E[T_x\mid\mathcal X]\).
Total variance gives \(\operatorname{Var}(T_{\mathcal X})\le
\operatorname{Var}(T_x)\).
The two events \(|Z_x-z_x|\le z_x/3\) and
\(|T_{\mathcal X}-t_x|\le t_x/3\) fail with probability at most
\(18V_0\). On their intersection the posterior mean under the uniform valid law is
at least \((t_x/z_x)/2\). Nonnegativity in \eqref{aff:posterior} and the
lower weight bound show that the selected posterior mean is at least
\(\ell(t_x/z_x)/2\).

For effective supports \(E,E'\), conditional covariance of their
row sign products, with \(\gamma_i\) the posterior mean \eqref{aff:posterior} of row \(i\), is exactly
\[
 \prod_{i\in E\mathbin\triangle E'}\gamma_i
 \left(1-\prod_{i\in E\cap E'}\gamma_i^2\right)\ge0.
\]
The upper weight bound therefore bounds the selected conditional variance
by \(\ell^{-2}\) times the conditional variance under the uniform valid law.
On the count event, \(Z_x\ge2z_x/3\), so its average is at most
\[
 \frac94\ell^{-2}\frac{\operatorname{Var}(T_x)}{z_x^2}
 \le\frac94\ell^{-2}(t_x/z_x)^2V_0.
\]
Conditional Chebyshev at deviation \(\ell(t_x/z_x)/4\) costs at most
\(36\ell^{-4}V_0\). Adding \(18V_0\) and the input-event failure
proves the assertion.
\end{proof}

In the application, \(\mathcal E\) can be taken to be the event
\eqref{low:energy-event}; intersect it with the count event already
included in the preceding proof. Equations~\eqref{low:population-counts}
and~\eqref{low:thinning-certificates} then imply the sampler certificates
\eqref{aff:certificates}. To formalize the conditional-variance argument,
define the input-only early law arbitrarily on inputs outside this
intersection; the proof uses its two-sided bounds only on the intersection.
With \(V_0\coloneqq4V_{\mathrm{guard}}\) and \(\ell\coloneqq\ell_0\), the error per
target is \(216\ell_0^{-4}V_{\mathrm{guard}}\), apart from the one
common event \eqref{low:energy-event}. The lower sign bound is at least
\(\ell_0\gamma_c(x)/16\), which implies the weaker
\(\ell_0\gamma_c(x)/32\) used in Section~\ref{sec:low-moment}.

\end{document}